\documentclass[11pt,a4paper]{article}
\usepackage{jcappub}
\usepackage{subfig}

\newcommand{\citeeq}[1]{Eq.~(\ref{#1})}

\newcommand{\citesec}[1]{Sect.~\ref{#1}}

\newcommand{\citetab}[1]{Tab.~\ref{#1}}
\newcommand{\citefig}[1]{Fig.~\ref{#1}}

\newcommand{\eg}{{\it e.g.}\ }
\newcommand{\ie}{{\it i.e.}\ }
\newcommand{\etc}{{\it etc.}\ }

\newcommand{\ben}{\begin{eqnarray}}
\newcommand{\een}{\end{eqnarray}}
\newcommand{\be}{\begin{equation}}
\newcommand{\ee}{\end{equation}}
\newcommand{\bei}{\begin{itemize}}
\newcommand{\ei}{\end{itemize}}
\newcommand{\besubeq}[1]{\begin{subequations} \label{#1}}
\newcommand{\eesubeq}{\end{subequations}}
\newcommand{\nn}{\nonumber}

\newcommand{\msun}{\mbox{$M_{\odot}$}}

\newcommand{\mchi}{\mbox{$m_{\chi}$}}
\newcommand{\mchisq}{\mbox{$m_{\chi}^2$}}

\newcommand{\sigv}{\mbox{$\langle \sigma v \rangle $}}
\newcommand{\rsun}{\mbox{$r_\odot$}}

\newcommand{\Nsub}{\mbox{$N_{\rm cl}$}}
\newcommand{\Nsubtot}{\mbox{$\bar{N}_{\rm cl}$}}

\newcommand{\fsub}{\mbox{$f_{\rm sub}$}}
\newcommand{\rhosub}{\mbox{$\rho_{\rm sub}$}}
\newcommand{\rhosm}{\mbox{$\rho_{\rm sm}$}}
\newcommand{\Msub}{\mbox{$M_{\rm cl}$}}

\newcommand{\MMW}{\mbox{$M_{\rm MW}$}}

\newcommand{\prob}[1]{\mbox{${\cal P}_{#1}$}}
\newcommand{\mymeanlr}[2]{\mbox{$ \left\langle {#1} \right\rangle_{#2} $}}

\graphicspath{{./Figures/}}

\subheader{LUPM:12-039}

\title{Diffuse gamma-ray constraints on dark matter revisited\\
  I: the impact of subhalos}

\author[a]{Steve Blanchet}
\author[b]{and Julien Lavalle}

\affiliation[a]{Institut de Th\'eorie des Ph\'enom\`enes Physiques, 
\'Ecole Polytechnique F\'ed\'erale de Lausanne, \\
CH-1015 Lausanne --- SWITZERLAND}

\affiliation[b]{Laboratoire Univers \& Particules de Montpellier [LUPM]
CNRS-IN2P3 / Universit\'e Montpellier II [UMR-5299]
Place Eug\`ene Bataillon - CC 72
F-34095 Montpellier Cedex 05 --- FRANCE}

\emailAdd{steve.blanchet@epfl.ch}
\emailAdd{lavalle@in2p3.fr}

\abstract{
We make a detailed analysis of the indirect diffuse gamma-ray signals from dark matter
annihilation in the Galaxy. We include the prompt emission, as well as the emission from inverse 
Compton scattering whenever the annihilation products contain light leptons. 
We consider both the contribution from the smooth dark matter halo and that from substructures. 
The main parameters for the latter are the mass function index and the minimal subhalo
mass. We use recent results from N-body simulations to set the most reasonable range of 
parameters, and find that the signal can be boosted by a factor ranging from 2 to 15 towards the
Galactic poles, slightly more towards the Galactic anticenter, with an important dependence on the 
subhalo mass index. This uncertainty is however much less than that of the extragalactic signal 
studied in the literature. We derive upper bounds on the dark matter annihilation cross section 
using the isotropic gamma-ray emission measured by Fermi-LAT, for two directions in the sky, the 
Galactic anticenter and the Galactic pole(s). The former represents the lowest irreducible signal 
from dark matter annihilation, and the latter is robust as the astrophysical background, dominated 
by the hadronic contribution, is rather well established in that direction. Finally, we show how 
the knowledge of the minimal subhalo mass, which formally depends on the dark matter particle 
interactions with normal matter, can be used to derive the mass function index.
}

\begin{document}
\maketitle

\section{Introduction}
\label{sec:intro}
It is now well established that the dominant component of matter in the universe is not conventional
baryonic matter, but dark matter (DM) 
\cite{1987ARA&A..25..425T,1991ApJ...376...51W,2011ApJS..192...18K}. Moreover, it 
is also known that DM should be relatively {\em cold}, \ie\ with a rather small free-streaming 
length at the matter domination period, in order to trigger the hierarchical formation of 
structures on time. The resulting cold dark matter (CDM) paradigm has been able to describe a wealth
of data amazingly well, from galactic to cosmological scales. Prime among candidates to 
explain where DM stems from is the {\em weakly interacting massive particle} (WIMP). Its popularity 
arises from the simple fact that assuming a new particle with weak interactions and a mass around 
the electroweak scale automatically leads to the correct relic abundance, while providing 
specific and potentially observable signatures \cite{1988ARNPS..38..751P,1996PhR...267..195J}. This 
non-trivial result, as well as theoretical motivations for having new physics at the weak scale, 
converge to make this scenario very appealing. Nevertheless, the CDM paradigm might still suffer 
from predicting too much power on small scales, but this remains an open issue 
\cite{2009NJPh...11j5029P}. In turn, such a small scale power could be exploited for discovery 
purposes \cite{1993ApJ...411..439S}, as we will discuss below.

DM candidates arising in particle theories aimed at solving specific theoretical issues of the
standard model (SM) are usually featured by the property of self-annihilation (or decay). This may 
be related to the stability of the proton, as is the case in the popular supersymmetric 
framework, and opens the possibility of detecting cosmic annihilation traces 
\cite{1978ApJ...223.1015G,1984PhRvL..53..624S,1985ApJ...299.1001K,1996PhR...267..195J}.
DM indirect detection has been extensively studied in the past ten years, 
with a boost of activity since the successful launch of the latest gamma-ray satellite, the 
Fermi-LAT, granting a much higher sensitivity to point sources, a better energy resolution, and an 
energy range extending to higher energies than its predecessors \cite{2009ApJ...697.1071A}. In 
particular, the sensitivity to DM annihilation signals is better than ever, and probing 
cross sections lying in the canonical range fixed by requiring the correct WIMP cosmological 
abundance is now possible. However, after over 3 years of data taking, no signal was found which 
could be unequivocally traced back to DM, aside from a potential line-like feature around 130 GeV 
very recently found in the diffuse emission around the Galactic center 
\cite{2012JCAP...07..054B,2012arXiv1204.2797W}. Since the latter remains to be unambiguously 
connected to DM, this conservatively implies that limits on DM model parameters could be extracted 
from the total observed gamma-ray flux. A non-exhaustive list of works using Fermi-LAT results is 
\cite{2009NuPhB.821..399C,2010NuPhB.831..178M,2010JCAP...03..014P,2010NuPhB.840..284C,2010JCAP...04..014A,2010JCAP...11..041A,2010JCAP...07..008H,2012PhRvD..85d3509A,2011JCAP...03..051C,2012JCAP...01..041A,2011arXiv1111.2835A}.
These studies differ in that the gamma-ray signal from DM annihilation was inferred either from the
Galactic, extragalactic, or both contributions. Moreover, different sets of data were considered, 
namely different directions in the sky or the isotropic diffuse gamma-ray background
\cite{2010PhRvL.104j1101A}. Here we would like to focus on the contribution coming from the 
Galactic halo only so as to minimize the theoretical uncertainties; we leave the extragalactic 
calculation for a future work. Note that, except for extreme astrophysical assumptions, the 
Galactic component usually provides more stringent constraints than the extragalactic one
\cite{2010JCAP...07..008H,2010JCAP...11..041A}.

In the works cited above, the treatment of the DM-induced photon emission typically 
suffers from the following shortcomings:
\bei
\item For final states involving charged leptons, the diffusion of subsequent electrons and 
positrons from the point of annihilation to the point where they produce gamma-rays by inverse 
Compton (IC) scattering is neglected.
\item DM substructures, which are expected to populate the Galactic halo in number, are 
often neglected, and, if not, mostly considered for the prompt emission; nevertheless, they are 
rarely taken into account for the IC contribution, for which spatial diffusion of electrons and 
positrons also plays an important role.
\ei
Regarding the first point, the diffusion of electrons is typically assumed to be negligible away 
from the Galactic center when computing the gamma-ray flux from DM annihilation. 
Intuitively, this might only be correct when the density gradient is small over large enough 
distances, larger than the typical diffusion length. This was verified to some extent in 
\cite{2011PhRvD..83b3518P} (see also \cite{2010NuPhB.831..178M}), but focusing mostly on regions 
where the locally constrained diffusion coefficient is still meaningful. This therefore mostly 
concerns the so-called diffusion zone. Nevertheless, DM annihilation also injects cosmic-ray (CR) 
electrons outside this region, which should also contribute to the overall IC gamma-ray production 
along the line of sight. Here, we will shortly discuss this statement further, and see to what 
extent the approximation of neglecting spatial diffusion holds. We will perform a full analysis 
of the transport issue in a subsequent paper.

As for subhalos, they are robustly predicted from structure formation, either in full 
analytic approaches of hierarchical clustering \cite{1974ApJ...187..425P,2001MNRAS.323....1S}, and 
in N-body numerical simulations (\eg~\cite{2008Natur.454..735D,2008ApJ...679.1260M,2008MNRAS.391.1685S}), although the resolution needed to probe the smallest scales within the WIMP paradigm,
down to $\lesssim 10^{-10}\,\msun$~\cite{2009NJPh...11j5027B}, is out of reach except at 
very large redshifts \cite{2005Natur.433..389D}. It was realized a long time ago that subhalos 
could play an important role in indirect detection at the Galactic scale by boosting the predicted 
DM signals~\cite{1993ApJ...411..439S,1999PhRvD..59d3506B,2003PhRvD..68j3003B,2007A&A...462..827L}. 
Their impact was also studied in early works on the DM-induced extragalactic gamma-ray emission and 
its possible detection~\cite{2002PhRvD..66l3502U}. Missing this part would have led to underpredict 
the extragalactic flux by a factor of at least $10^4$. Consistency would therefore imply to 
systematically include DM substructures in indirect detection calculations. However, since the Fermi
data were released, the tendency in studies on DM indirect detection with diffuse gamma rays has 
often been either to neglect them or to deal with them with very crude estimates (except in a few 
cases, \eg~ \cite{2011PhRvD..83b3518P,2010JCAP...11..041A,2010JCAP...04..014A}).

Within the CDM paradigm, our Galactic halo is big enough to host a large number of DM 
clumps. N-body {\em zoomed-in} simulations have already resolved tens of thousands of them in 
Milky-Way-like halos down to their resolution limit ($\sim 10^4\,\msun$), and a conservative 
extrapolation to Earth-mass objects would even imply a total number of the order of, or larger than 
10$^{13}$. The presence of such Galactic subhalos must increase the diffuse Galactic gamma-ray
emission, especially at large latitudes. Although the amplification is expected to be modest, less 
than a factor of 10 typically \cite{2011PhRvD..83b3518P}, we emphasize that given the current 
constraints on small WIMP masses around 10 GeV or so, a factor of a few may have an important 
impact on still allowed borderline configurations. Moreover, we stress the theoretical 
uncertainties due to subhalos are much lower at the Galactic scale than at the cosmological 
scale~\cite{2002PhRvD..66l3502U,2011PhRvD..83b3518P,2010JCAP...04..014A}, the latter being relevant
to the extragalactic component; limits coming from predictions focused on the Galactic scale only
can therefore be considered as more robust. Nevertheless, in order to have a realistic 
prediction for the Galactic DM-induced gamma-ray flux on Earth, it is crucial to include subhalos 
in a consistent way, namely for both the prompt and IC signals. This is another goal of ours for 
this paper.

It is important to discuss which data sets and directions in the sky should be preferred in order to
derive robust bounds on the annihilation cross section. Clearly, the total gamma-ray flux measured 
by Fermi-LAT in every direction in the sky is a possible choice, but it is most probably too 
conservative. Indeed the conventional Galactic diffuse component induced by astrophysical CRs makes 
up a prominent part (if not most) of the total measured flux. The Fermi collaboration 
has actually recently released an extensive work addressing the CR-induced Galactic 
diffuse emission in detail~\cite{2012ApJ...750....3A}. We note that the main 
contribution comes from hadronic processes involving CR nuclei (mostly protons) and the interstellar
gas, while a subdominant one comes from the IC scattering of CR electrons off
the interstellar photon field (including the cosmic microwave background --- CMB). Besides,
the former is subject to less uncertainty than the latter. Moreover, we have a large number of 
(mostly extragalactic) astrophysical sources (\eg\ blazars or star-forming galaxies) which feature 
a substantial fraction of the total flux. Subtracting the Galactic diffuse emission and the emission
from known sources from the total flux was one of the many important contributions by the Fermi 
collaboration~\cite{2010PhRvL.104j1101A}, leading to a residual component found 
isotropic, the so-called diffuse isotropic gamma-ray background (IGRB). In this work,
we will use this estimate of the IGRB to set limits on DM annihilation cross-sections. Although
doing so exposes us to the modeling used by the Fermi collaboration to subtract the Galactic 
diffuse emission, we are confident that in the direction of the Galactic poles (GPs), the IGRB 
offers a robust upper limit for any gamma-ray flux from DM annihilation. We will also calculate 
fluxes in the direction of the Galactic anticenter (GAC), because it is the smallest possible 
gamma-ray flux and therefore acts as an irreducible isotropic component. However, in this direction,
the subtraction of the Galactic diffuse emission by the Fermi collaboration is more uncertain, as 
we will discuss.

This paper is organized as follows. In \citesec{sec:astro} we introduce the main definitions, and 
provide some important astrophysical parameters that will be used in the remainder of the paper. 
In \citesec{sec:gamma}, we present the different sources of gamma-rays stemming from DM annihilation
(prompt and IC), and provide the formul\ae\ used to compute gamma-ray fluxes on Earth 
in any direction in the sky. \citesec{sec:transport} is devoted to the transport of electrons and 
positrons in the Galaxy. \citesec{sec:sub} shows how we include Galactic subhalos
within the calculation of the diffuse gamma-ray fluxes. We finally present our main results in
\citesec{sec:res}, and draw our conclusions in \citesec{sec:concl}.

%
\section{Astrophysical parameters: DM and target radiation fields in the Galaxy}
\label{sec:astro}
The DM halo of our Milky Way (MW) may be well described by the so-called Navarro-Frenk-White (NFW) 
profile~\cite{1997ApJ...490..493N}
\ben
\label{eq:densprof}
\rho(r)={\rho_s \over r/r_s (1+r/r_s)^2} \, ,
\een
where $r_s$ is the scale radius, and $\rho_s$ is the scale density. In the following,
we will use the parameters determined in a recent study of the Galactic kinematic data 
\cite{2011MNRAS.414.2446M}, namely a scale radius of $20.2$~kpc, a local density of 
$\rho_{\odot}=0.395$~GeV/cm$^3$, and a Sun-GC distance of $r_{\odot}=8.29$~kpc, which, put together, 
fix the scale density given above. These parameters are in agreement with those found in other 
complementary works \cite{2010JCAP...08..004C,2010A&A...523A..83S,2012arXiv1205.4033B}.

The interstellar radiation field (ISRF), which defines the target photons for IC scattering, is 
made of three main components \cite{2005ICRC....4...77P}: starlight (SL), infrared (IR) light 
emitted by the interstellar gas, and the CMB. We assume these components to obey blackbody 
distributions such that the differential number density for any kind $i$ reads:
\ben
\label{eq:isrf}
{{\rm d} n_i \over {\rm d} E}(E,r',z)= 
n^0_{i} \, f(r',z) {E^2\over \pi^2} {1\over \exp (E/T_i) -1} \, ,
\hspace{1cm} i={\rm SL}, {\rm IR}, {\rm CMB}
\een
where $(r',z)$ are cylindrical coordinates centered on the Sun's position, and, except
for the CMB for which $f(r',z)=1$, we follow Ref. \cite{2011ApJ...727...38S} for the 
parameterization of the position dependent normalization:
\ben
\label{eq:fisrf}
f(r', z)= \exp\left[-\left(r'/r_{\rm ph}+|z|/z_{\rm ph}\right)\right] \, ,
\een
with $r_{\rm ph}=3.2$~kpc and $z_{\rm ph}=0.4$~kpc. The normalizations as well as
reference temperatures $T_i$ are given in \citetab{tab:table1}.
\begin{table}
\centering
  \begin{tabular}{l | c c c}
    & CMB & IR & SL \\ \hline
    $T_i$~[eV] & $2.35\times 10^{-4}$ & $2.85\times 10^{-3}$  & $2.8\times 10^{-1}$ \\ \hline
    $n_i^0/n_i^{\rm bb}$ & 1 & $4.5\times 10^{-5}$ & $7\times 10^{-13}$ \\ \hline
  \end{tabular}
\caption{ISRF temperatures and density normalizations with respect to the blackbody reference 
  values.}
\label{tab:table1}
\end{table}

We further account for the presence of a magnetic field in the Galaxy which enters the electron 
energy losses (see \citesec{subsec:losses}), the energy density of which can be expressed as:
\ben
\label{eq:ub}
u_B(r,r)&=& {B^2(r,z) \over 8 \pi}\, ,\\
B(r,z) &=& B_\odot \, \exp\left(-{(r-\rsun)\over r_B}+{|z|\over z_B}\right)\, .\nn
\een
We normalize the magnetic field to a local value of $B_\odot = B(r=\rsun,z=0) = 3~\mu$G. This 
corresponds to a magnetic field of $\sim 7~\mu$G at the GC. Following \cite{2000ApJ...537..763S} 
we choose typical values of $r_B= 10$~kpc and $z_B=2$~kpc.

%
%
\section{Gamma-ray emission}
\label{sec:gamma}
\subsection{Prompt gamma-ray contribution}
\label{subsec:prompt}
If DM annihilates into charged particles, photons may be emitted in two ways, which we collectively 
refer to as {\em prompt} emission:
(i) if the final state particles hadronize, they will produce neutral pions which decay
into photons; (ii) charged particles will radiate photons from internal Bremsstrahlung processes 
\cite{2005PhRvL..94q1301B}. In the following, we will assume that DM is made of Majorana 
particles, denoted $\chi$. The prompt photon flux collected on Earth from DM annihilation in the 
Galactic halo, along a line of sight (los) $ds$ and within a solid angle $\delta\Omega_{\rm res}$ is 
given by
\ben
\label{eq:prompt}
{ {\rm d}\Phi \over {\rm d} E_{\gamma}} = 
{r_{\odot} \over 4\pi}\, \left\{ {\cal S}\equiv {\sigv  \over 2}  \,
{\rho_{\odot}^2 \over \mchisq} \right\}  \,
{{\rm d} N_{\gamma}\over {\rm d} E_{\gamma}} \int_{\delta\Omega_{\rm res}} {\rm d}\Omega
\int_{\rm los} {{\rm d}s\over r_{\odot}} \left({\rho(r)\over \rho_{\odot}}\right)^2 \, ,
\een
%
where ${\rm d}N_{\gamma}/{\rm d}E_{\gamma}$ is the photon
spectrum per annihilation, $\sigv$ is the thermally averaged annihilation cross-section, and the 
density profile is given in \citeeq{eq:densprof}. The galactocentric radius $r$ can be expressed 
in terms of the line-of-sight distance $s$ and the galactic coordinates $(b,\ell)$:
\ben
\label{eq:rlos}
r(s)=\sqrt{r_{\odot}^2- 2\,s\,r_{\odot}\,\cos(\ell)\,\cos(b) +s^2}\, .
\een
\subsection{Inverse Compton scattering}
\label{subsec:ics}
The most recent analyses focused on DM models which couple dominantly to the lepton 
sector (see~\eg~\cite{2009JCAP...02..021I,2010JCAP...01..009I,2010NuPhB.831..178M,2010JCAP...03..014P,2010JCAP...07..008H,2012PhRvD..85b3004C}). Such models may account for the rising positron
fraction observed by the PAMELA and Fermi 
experiments~\cite{2009Natur.458..607A,2012PhRvL.108a1103A}, and in some 
cases the electron-positron spectral feature found around a few hundreds of GeV with the Fermi-LAT 
instrument~\cite{2009PhRvL.102r1101A}.
However, it must be kept in mind that such scenarios will typically produce an overabundance of 
photons as well, due to IC scatterings from final-state electrons and 
positrons~\footnote{For our purposes, the fact that electrons and positrons have a different charge 
has no importance, and therefore, from now on, we will refer to {\em electrons and positrons} as 
simply {\em electrons}.}. It is then a quantitative question whether such models are excluded or 
not, and many groups have contributed to this discussion
\cite{2010NuPhB.831..178M,2010JCAP...03..014P,2010JCAP...11..041A,2010JCAP...07..008H,2010NuPhB.840..284C,2012PhRvD..85d3509A}.

The photon energy spectrum arising from the IC scattering of an incoming 
electron of energy $E$ (and Lorentz factor $\gamma_e=E/m_e$) off an incoming photon of energy 
$E_{\rm in}$ is given by~\cite{1965PhRv..137.1306J,1968PhRv..167.1159J,1970RvMP...42..237B}
\ben
\label{eq:power}
{\cal P}_i(E_{\gamma}, E, \vec{x}) &=& 
{3 \sigma_T \, c \over 4 \gamma_e^2} \int_{1/4\gamma_e^2}^1 {\rm d} q \,
{1\over q}\,{{\rm d} n_i \over {\rm d} E_{\rm in}}(E_{\rm in}(q),\vec{x}) \, f(q) \, ,\\
f(q) &=& 1 + 2q \left\{ \log q - q + {1\over 2} \right\}  + 
{(1-q)\over 2}{(\Gamma_{e}q)^2\over 1+\Gamma_{e}q}\, ,\nn
\een
where $q={E_{\gamma}\over \Gamma_{e}(E- E_{\gamma})}$, $\sigma_T = 0.665$~barn is the 
Thomson cross section, $\Gamma_{e}=4E_{\rm in}\gamma_e/m_e$ is a dimensionless parameter that 
determines the regime of the scattering, \eg\ the Thomson limit when $\Gamma_{e}\ll 1$. 
We show in \citefig{fig:PhotonPower} how this function depends on the energy of the outgoing 
photon $E_{\gamma}$, and on the incoming electron energy $E$. As explained in \citesec{sec:astro},
the target photons here can be the CMB, IR light or starlight, represented by the 
subscript $i$ in \citeeq{eq:power}. In \citefig{fig:PhotonPower}, where we show the total photon 
spectrum for incoming electrons of different energies, 
these three components appear as bumps, the 
leftmost one originating from electrons scattering off the CMB, the middle one off IR, and the 
rightmost one off SL. These bumps unveil the initial target photon distributions (taken
thermal here), and feature the original peaks, initially around $E_{\rm in} \approx T_i$, which have
been IC-upscattered to energy $E_\gamma \approx \gamma_e^2\, T_i$.
\begin{figure}
  \begin{center}
    \input{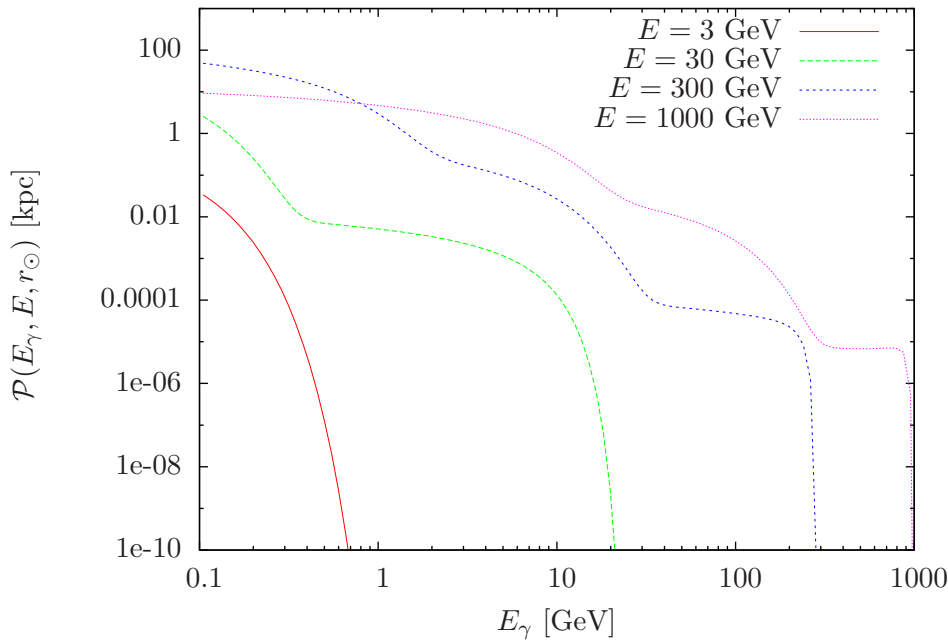}
  \end{center}
  \caption{Differential photon power emitted at $r=r_{\odot}$ by an incoming electron of energy $E$.} 
  \label{fig:PhotonPower}
\end{figure}
One can then express the photon flux on Earth due to such processes, as observed in a solid angle
$\delta \Omega_{\rm res}$, given a differential electron number density 
${\cal N}_e (\vec{x}, E)\equiv d n_e / d E$:
\ben
\label{eq:fluxIC}
{ {\rm d}\Phi \over {\rm d} E_{\gamma}} = { r_{\odot}  \over 4\pi} \,
\int_{\delta \Omega_{\rm res}} {\rm d}\Omega \int_{\rm los} {{\rm d}s\over r_{\odot}}  
\int_{m_e}^{\mchi} {\rm d}E \, {\cal N}_e (\vec{x}(s,\theta,\phi), E)  \, 
\sum_i {\cal P}_i(E_{\gamma}, E, \vec{x}(s,,\theta,\phi)) \, .
\een
In order to calculate the photon flux from IC scattering, one therefore needs to know the electron 
number density produced from DM annihilation along the line of sight $s$. We will show how to 
calculate this density everywhere in the galaxy in \citesec{sec:transport}, taking into account 
spatial diffusion effects. We note that \citeeq{eq:fluxIC} implicitly assumes that the incoming 
electron and photon fluxes are isotropic at position $\vec{x}$, which should be questioned in the 
present context. We will shortly discuss this in \citesec{sec:transport}, but we will dedicate a 
more complete study of this issue in a forthcoming paper.
\begin{figure}
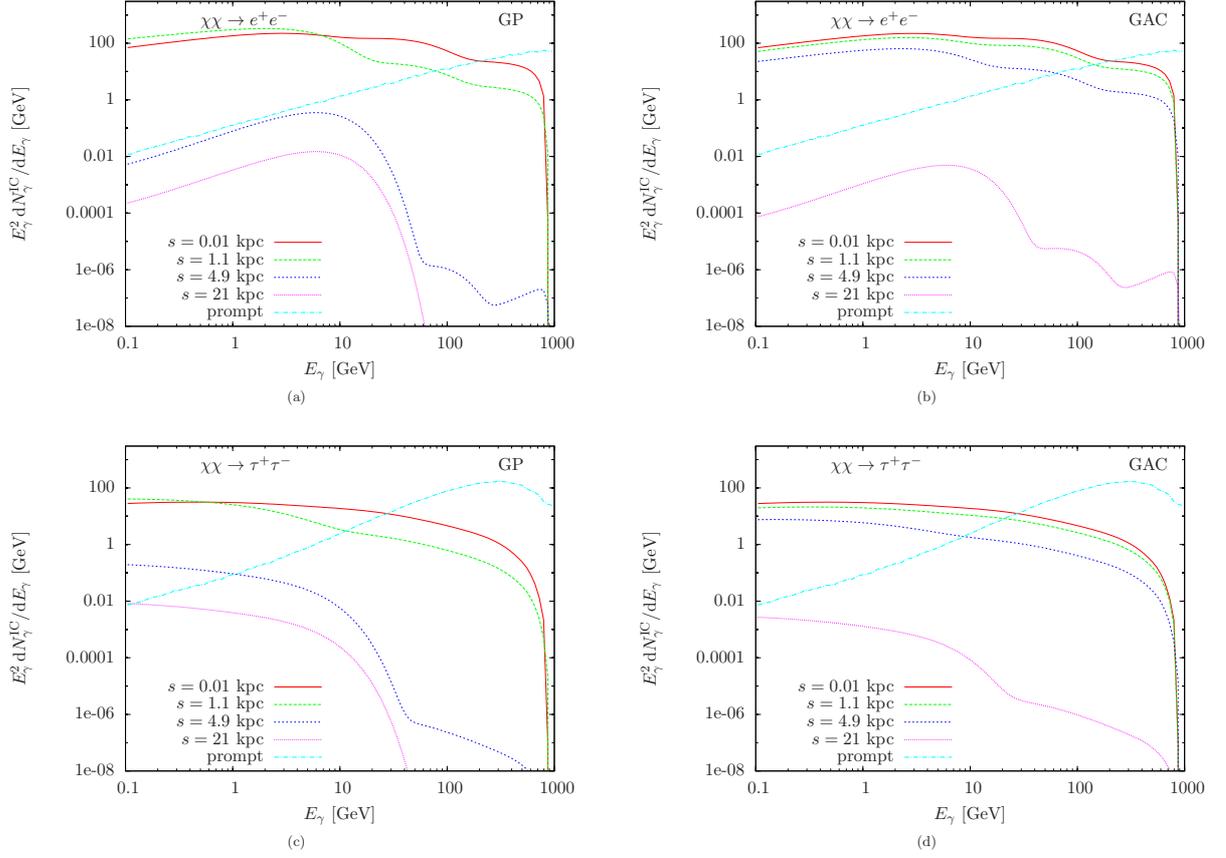

\scalebox{0.6}{\subfloat[]{\input{Figures/ICSpecHLLe_989.tex}}}
\hspace{0.5cm}
\scalebox{0.6}{\subfloat[]{\input{Figures/ICSpecAGCe_989.tex}}}
\vspace{0.1cm}
\scalebox{0.6}{\subfloat[]{\input{Figures/ICSpecHLLtau_989.tex}}}
\hspace{0.5cm}
\scalebox{0.6}{\subfloat[]{\input{Figures/ICSpecAGCtau_989.tex}}}
  \caption{Differential photon spectrum produced for a 1 TeV DM particle annihilating into $e^+e^-$ 
    (top) and $\tau^+\tau^-$ (bottom). The IC contribution is shown at different distances on the 
    l.o.s. towards the GP (left) and the GAC (right) and the prompt contribution is shown as a 
    reference.} 
  \label{fig:PhotonSpec}
\end{figure} 

We show in \citefig{fig:PhotonSpec} the prompt/IC spectra associated with different masses of WIMPs 
annihilating into $e^+e^-$ for two different lines of sight: the Galactic anticenter (GAC) and the 
Galactic pole(s) (GP). These spectra correspond to the energy integral of \citeeq{eq:fluxIC},
normalized to the local annihilation rate ${\cal S}$ defined in \citeeq{eq:prompt}: 
\ben
\label{eq:specIC}
{ {\rm d}N^{\rm IC}_{\gamma}(s,b,\ell) \over {\rm d} E_{\gamma}} = {1\over {\cal S}}  
\int_{m_e}^{\mchi} {\rm d}E \,
\, {\cal N}_e (\vec{x}(s,b,\ell), E)  \, \sum_i {\cal P}_i(E_{\gamma}, E, \vec{x}(s,b,\ell)) \, .
\een
It can 
be seen that the spectrum drops quite dramatically at high energies for the GP because the target 
photon density decreases fast as one goes away from the Galactic disk (see next section). The CMB 
being present everywhere, its bump is only affected by the smaller energy losses away from the 
Galactic center. Increasing the distance towards the GAC also shows a decrease in the high-energy 
spectrum, but to a much lesser extent than the GP. One can also notice that the prompt component 
starts dominating at the high-energy end of the spectrum, as expected. DM annihilation into 
$\tau^+\tau^-$ exhibits the same features. The main differences are:
1) the IC spectrum is shifted to lower energies and {\em smeared} because the electron energy 
distribution is not as sharp as the $e^+ e^-$ case; 2) the prompt component is more dominant here 
because $\tau$ can decay hadronically producing many more photons. We have made use of the 
Pythia Monte Carlo generator \cite{2006JHEP...05..026S} to compute the injected electron spectra 
in all cases.

\section{Transport of Galactic electrons}
\label{sec:transport}
\subsection{Generalities}
\label{subsec:gen}
Electrons produced from DM annihilation may scatter off Galactic magnetic inhomogeneities, which 
induces a diffusive motion, and lose energy mostly through synchrotron and inverse Compton (IC) 
processes in the energy range considered here. These losses give rise to a broad 
electromagnetic spectrum that may help, in turn, trace or constrain the electron distribution from 
radio frequencies (synchrotron) to gamma-rays (IC), which we are interested in in this paper. In 
order to calculate the differential electron number density at all times and positions, one needs to
solve a diffusion-loss equation~\cite{1959SvA.....3...22S,1964ocr..book.....G,berezinsky_book_90}. 
Neglecting convection and reacceleration, which become sizable only below a few GeV 
\cite{2009A&A...501..821D}, the diffusion-loss equation for electrons reads
\ben
\label{eq:diffeq}
\partial_t \, {\cal N}(E,\vec{x}) - 
\vec{\nabla}\, \left\{K(E,\vec{x}) \,\vec{\nabla} \,{\cal N}(E,\vec{x}) \right\}
- \partial_E \left\{ b(E,\vec{x}) {\cal N}(E,\vec{x}) \right\} = {\cal Q}(E,\vec{x}) \, ,
\een
where ${\cal Q}$ is the electron source term, and diffusion off magnetic turbulences is described 
with a diffusion coefficient $K(E,\vec{x})$ which is energy and {\em a priori} spatial dependent.
The energy losses $b(E,\vec{x}) = -dE/dt$ are also spatial dependent since, beside including the 
CMB, they are characterized by interactions with the magnetic field and the interstellar radiation 
field --- energy losses are discussed into more detail in \citesec{subsec:losses}. In this paper, 
the source term is proportional to the squared DM density (assuming Majorana WIMPs):
\ben
\label{eq:source}
Q(E_s,\vec{x}_s)= {\cal S} \, 
\left[ {\rho(\vec{x}_s) \over \rho_\odot } \right]^2 \, {{\rm d}N_e\over {\rm d} E_s}  \, ,
\een
where the local annihilation rate ${\cal S}$ was defined in \citeeq{eq:prompt}. 
We note that when subhalos are included --- see \citesec{sec:sub} --- the source can be
split into two separate terms~\cite{2007A&A...462..827L}. The first one regards the smooth halo 
component, and corresponds to \citeeq{eq:source} with the substitution $\rho \to \rhosm$. The 
second one, which will be fully derived in \citesec{sec:sub} (all relevant definitions will be found
there), reads
\ben
Q_{\rm sub}(E_s,\vec{x}_s)= 
{\cal S} \, \Nsubtot \, \mymeanlr{\xi(r)}{M} \,
{ d\prob{V}(r) \over d^3\vec{x}_s} \, {{\rm d}N_e\over {\rm d} E_s}  \, .
\een
Since the source is stable over cosmological times, one can safely take the steady state limit of 
\citeeq{eq:diffeq}, which can be solved either numerically 
(\eg~\cite{1998ApJ...509..212S,2009arXiv0909.4548D,2011JCAP...03..051C}), or, given some 
approximations, with semi-analytic methods (\eg~\cite{1974Ap&SS..29..305B,1998PhRvD..59b3511B,2007A&A...462..827L,2011ApJ...727...38S}).

In the following, we will discuss potential ways to solve \citeeq{eq:diffeq} which may arise
in the context of DM annihilation. To do so, we may assume that any solution can 
be expressed in the form of a Green function ${\cal G}$ such that the differential electron 
density ${\cal N}$ can be calculated from
\ben
\label{eq:psiGreen}
{\cal N}(E,\vec{x})= \int_{E_s=E}^{E_s=\infty} {\rm d} E_s 
\int {\rm d}^3 \vec{x}_s \,{\cal G} ( E, \vec{x}  \leftarrow E_s , \vec{x}_s)  \, 
Q(\vec{x}_s,E_s) \, .
\een
Such a Green function can then be interpreted as the probability for an electron injected at 
position $\vec{x}_s$ with energy $E_s$ to diffuse to another position $\vec{x}$ down to energy 
$E<E_s$.

\subsection{Electron transport models for DM-induced IC calculations}
\label{subsec:transp_models}

When energy losses and the diffusion coefficient are taken spatially homogeneous, and when 3D 
spatial boundaries are pushed away towards infinity, then the Green function takes a rather simple 
expression \cite{1959SvA.....3...22S}, which will turn to be useful later on:
\ben
\label{eq:green3D}
{\cal G}(E, \vec{x} \leftarrow E_s,\vec{x}_s) &=& {1 \over b(E) \, (\pi \,\lambda^2)^{3/2} } \exp 
\left\{- {| \vec{x}-\vec{x}_s |^2 \over \lambda^2 } \right\}\, , \\
\text{where} \;\; \lambda^2(E,E_s)&=&
4 \int_E^{E_s} {\rm d} E' {K(E') \over  b(E')} \, .
\label{eq:lambda}
\een
where $\lambda$, a function of energy, can be interpreted as the electron propagation length, and
carries the dimension of a distance; it is typically of the kpc order on average in the MW. Note 
that when cylindrical boundary conditions are implemented, semi-analytic solutions to 
\citeeq{eq:diffeq} do exist still in the form of Green functions or Fourier-Bessel expansions 
(\eg~\cite{1974Ap&SS..29..305B,1998PhRvD..59b3511B,2007A&A...462..827L,2008PhRvD..77f3527D,
2012arXiv1205.1004L}). When parts of the energy losses and sources can be assumed as fully confined 
to the disk, some tricks can be used to accurately predict the local density of electrons 
{\em at the Earth} \cite{2009A&A...501..821D,2011ApJ...727...38S}. When energy losses have a more 
complicated spatial distribution, then it is difficult to avoid a full numerical treatment, unless 
specific energy range or spatial distributions of the relevant ingredients (including the source) 
are considered (\eg~\cite{2011ApJ...727...38S}). In this paper, we will slightly refine the modeling
already presented in \cite{2010A&A...524A..51D}.

The main difficulty arising when trying to predict the IC contribution to the diffuse 
gamma-ray flux is to estimate the electron distribution at each point along the line of sight,
either inside the magnetic halo where diffusive motion is valid, and outside where
\citeeq{eq:diffeq} is no longer valid but DM annihilation still takes place --- another
issue is related to the spatial dependence of the electron energy losses, which we will
discuss later on. This defines two main sources of uncertainty: (i) the size of the magnetic halo 
(essentially the vertical extent), and (ii) the treatment of CR transport outside it.

Most propagation models are based on the assumption that the CR density vanishes outside the 
magnetic halo. This is actually an empirical way to feature the transition between diffusion and 
ballistic motion, which is quite appropriate for standard CRs which are either primaries accelerated
at astrophysical sources or secondaries stemming from spallation processes, both injected in the 
Galactic disk: those CRs may diffuse in an extended magnetic halo, but when they reach the spatial 
boundaries, they cannot, in principle, be scattered backward anymore by magnetic inhomogeneities, 
which should have also vanished. This process should in principle be described by relating the 
diffusion coefficient to the spatial and spectral properties of the magnetic turbulence without 
imposing any fixed spatial boundary~\cite{2002PhRvD..65b3002C,2009ncrd.book.....S}, but (a) such a 
refined modeling would be hard to constrain from observations, and (b) the underlying 
phenomenology is not fully established yet. Instead, a much more simple picture is widely used, 
which is phenomenologically similar, in which the diffusion coefficient is taken homogeneous inside 
an extended cylinder encompassing the Galactic disk, beyond which the CR density is assumed to 
vanish, which amounts to imposing spatial boundaries to \citeeq{eq:diffeq}. Cylindrical symmetry 
makes the calculations much simpler.

{\bf Inside the magnetic halo:}
The impact of the uncertainties in the half-width of the magnetic halo, $L$, on the DM-induced 
CR flux or density has been widely investigated in the past 
(\eg~\cite{2004PhRvD..69f3501D,2008A&A...479..427L,2008PhRvD..77f3527D,2010PhRvD..82h1302L}),
and can be featured by rather large theoretical errors. Nevertheless, most studies have 
considered a quite extreme and {\em ad hoc} range for $L$, 1-15 kpc, originally proposed in 
\cite{2004PhRvD..69f3501D} and conservatively motivated by the blind analysis performed in 
\cite{2001ApJ...555..585M} more than a decade ago. We emphasize that the status of small
values for $L$ has been recently considerably revised, and it is generally expected that
$\gtrsim 3$-$4$ kpc 
\cite{2010A&A...516A..66P,2009PhRvL.103y1101A,2010PhRvD..82h1302L,2010PhRvL.104j1101A,2012ApJ...750....3A}. Beside the weak constraint coming from radioactive CR species 
\cite{1998ApJ...509..212S,2010A&A...516A..66P}, a strong argument against a small $L$ comes from a 
conflict with the CR positron data \cite{2011MNRAS.414..985L}. Likewise, diffuse gamma-ray emission 
models also suggest large values for $L$ to enhance the relative Galactic IC contribution and
improve the global fit \cite{2012ApJ...750....3A}. With such intermediate to large values for $L$, 
the theoretical uncertainties in the predictions decrease significantly and are therefore less 
concerning than before.

{\bf Outside the magnetic halo:}
As to the treatment of the transport outside the magnetic halo, it has often been modeled
from some misconceptions. Many authors have argued that spatial diffusion could be neglected
outside the magnetic halo, and solved \citeeq{eq:diffeq} by dropping the diffusion term; some others
adopted this assumption mostly in the absence of a complete and self-consistent transport model 
able to describe the transition between diffusive and ballistic motion (\eg\ 
\cite{2009NuPhB.821..399C,2010NuPhB.831..178M,2010NuPhB.840..284C,2010PhRvD..81j3521K,2010JCAP...07..008H}). We emphasize that independently of the motivation, this is erroneous. Indeed,
neglecting spatial diffusion in \citeeq{eq:diffeq} amounts to assuming that electrons
lose all their energy before propagating over significant distances. This would be correct
only if the source did not exhibit any spatial gradient over such distances. This is not the case,
since outside the diffusion zone, energy losses are much less than important inside (only CMB is 
present), while the diffusion coefficient should be much larger (it should formally tend to 
infinity). Hence, the propagation scale given in \citeeq{eq:lambda} should tend to very large 
values, which can even exceed the DM halo size (see also \cite{2010PhRvD..82h3521L}). This should 
translate into a complete dilution of the electron density. In contrast, neglecting spatial 
diffusion in \citeeq{eq:diffeq} leads to a very large overestimate of the electron density 
outside the magnetic halo.

Nevertheless, irrespective of the diffusion coefficient value, there should still be an asymptotic
case for which the propagation length $\lambda \to 0$ even outside the magnetic halo (say even in 
the limit of $K\to\infty )$. This actually occurs when the electron energy tends to its injected 
value. Even when $K$ is large, $\lambda \overset{E\to E_s}{\longrightarrow} 0$. This helps figure 
out a phenomenological strategy to describe transport from the border to outside the diffusion zone.
We may neglect contributions for which the energy-loss timescale is sufficient for the electron 
density to be diluted significantly and plug the following Green function:
\ben
\label{eq:gnodiffb}
{\cal G}_{E\to E_s} (E,\vec{x}\leftarrow E_s ,\vec{x}_s) = {\delta^3(\vec{x}-\vec{x}_s) 
\, \theta(E+\delta E-E_s) \over b(E,\vec{x})}
\,,
\een
where $\delta E \to 0$, and is fixed, in practice, such that $c\int_E^{E+\delta E} dE'/(b(E',\vec{x}) 
< 1$ kpc. We note that such an asymptotic regime is actually also valid in the confinment zone,
so the spatial dependence of the energy losses can be safely taken into account as long as we are
close to the limit $\lambda \to 0$. This induces an important normalization effect, especially
at high energy where $\lambda \to 0$ quite generically, since $b(E)$ is much larger in the Galactic 
disk than away from it, where the only target photon field is the CMB. This will discussed into
more detail below.

Actually, CR transport outside the diffusion zone has already been investigated 
(see \eg~\cite{2002A&A...388..676B,2010PhRvD..82d3505P,2011PhRvD..83l3508P}). 
Refs. \cite{2002A&A...388..676B,2011PhRvD..83l3508P} treat the antiproton case, for which energy
losses are almost irrelevant, which makes it difficult to compare with electrons.
On the other hand, Ref. \cite{2010PhRvD..82d3505P} addresses the positron case with a 3-zone 
propagation model, where the usual diffusion zone is embedded into a vertically more extended 
region characterized by a much larger diffusion coefficient. The impact on the 
diffuse gamma-ray emission was found to be small. Nevertheless, energy losses 
were taken homogeneous, which may induce some discrepancy in the high-energy limit of the electron 
density along the line of sight, as discussed below \citeeq{eq:gnodiffb}. Typically, the electron
density should scale linearly with the energy loss timescale ($\propto 1/b(E,\vec{x})$), which is 
expected to increase with the latitudinal distance until saturating to the CMB value (decrease of 
the magnetic field and ISRF amplitudes down to zero): a transport model with homogeneous losses 
normalized to the solar system's values does underpredict the asymptotic high energy electron 
density, and does therefore underpredict the associated IC emission. To circumvent this potential 
issue, we will account for the spatial dependence of the energy-loss term in an effective manner
by using an average value $\langle b(E,\vec{x})\rangle_\lambda$ such that 
$\langle b(E,\vec{x})\rangle_\lambda \overset{\lambda \to 0}{\longrightarrow} b(E,\vec{x})$ is 
ensured for small values of the propagation scale $\lambda$. All this is illustrated in 
\citefig{fig:comp_prop_mod}, where we compute the IC gamma-ray flux associated with the 
annihilation channel $\chi\chi\to e^+ e^-$, with $\mchi = 1$ TeV --- we assume the DM to be 
smoothly distributed according to the NFW profile given in \citeeq{eq:densprof}, and take a Dirac
function for the injection spectrum, in contrast to the final results that will be derived
by using a full Pythia \cite{2006JHEP...05..026S} spectrum (which is slighly broader than the Dirac 
function due to Bremsstrahlung effects). The solid blue (or dark) line is the IC flux predicted 
when the CMB is taken as the unique ISRF component, the dotted line corresponds to a full account 
of the local ISRF components assumed homogeneous, while the dashed line and the red (or gray) solid 
line encode the inhomogeneous distribution of the ISRF. We see that in the second case, the IC flux 
is significantly underpredicted, as emphasized above. Alternatively to the method presented here, 
note that a proper treatment of spatial-dependent energy losses is ensured when \citeeq{eq:diffeq} 
is solved numerically \cite{1998ApJ...509..212S,2009arXiv0909.4548D,2011JCAP...03..051C}.
\begin{figure}
\centering
\includegraphics[width=0.6\textwidth]{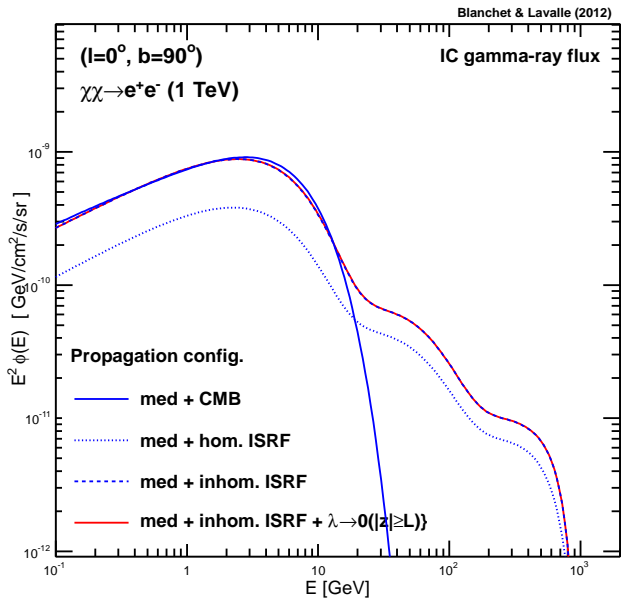}
\caption{IC gamma-ray flux assuming annihilation into $e^+e^-$ (1 TeV). The {\em med} set of
propagation parameters has been adopted \cite{2001ApJ...555..585M,2004PhRvD..69f3501D}.
The solid blue (dark) curve shows the prediction when the CMB is the only ISRF component; the dotted
curve corresponds to the prediction for which all ISRF components are included but taken 
homogeneous in the diffusion zone; the dashed curve accounts for the spatial distribution of the
ISRF; the solid red (gray) curve promotes the $E \to E_s$ regime as valid beyond the vertical 
boundary.}
\label{fig:comp_prop_mod}
\end{figure}

For completeness, we may mention a last source of theoretical uncertainty which comes from the 
angular dependence of the IC scattering cross section. Indeed, the IC flux expression given in 
\citeeq{eq:fluxIC} implicitly assumes that the incident electrons have an isotropic distribution of 
momenta~\cite{1965PhRv..137.1306J,1968PhRv..167.1159J,1970RvMP...42..237B}. Nevertheless, we may 
expect that when magnetic inhomogeneities get scarce, electrons are no longer isotropized, and the 
flux escaping from central Galactic regions is larger than the flux coming from opposite directions,
as it is the case for gamma-rays. Assuming the electron flux is isotropic outside the diffusion 
zone will therefore likely give rise to an overestimate of the smooth halo contribution to the IC 
flux, since the IC scattering cross section formally depends on the scattering angle ---
the emitted photon is collimated along the initial electron momentum \cite{1981Ap&SS..79..321A}. 
This angular effect mostly regards the smooth DM contribution, since subhalos are DM overdensities 
encountered along the line of sight\footnotemark. We note that one recovers isotropy in the 
asymptotic regime described by \citeeq{eq:gnodiffb}, which involves electrons produced locally only.
Since we will use the latter regime outside the diffusion zone, our calculations will be
conservative in this respect.

\footnotetext{It is interesting to note that this anisotropy is not only characteristic of the
electron flux outside the diffusion zone, but also of the ISRF fields in the disk --- this
is particularly relevant to compute the IC gamma-ray flux when the line of sight goes along the
disk \cite{2000ApJ...528..357M}.}

\textbf{Summary:}
In the following, we will use the slab model with usual boundary conditions except in the limit 
$\lambda \overset{E\to E_s}{\longrightarrow} 0$, where we promote continuity with the 
outside regions by means of the asymptotic 3D solution of \citeeq{eq:gnodiffb}. In practice, we 
use a slightly modified slab model in the sense that we use it with the 
{\em locally averaged} value of the energy losses along the line of sight. This approximation 
ensures the correct asymptotic value of the electron density everywhere in the halo when 
$\lambda\to 0$, which is critical in the IC calculation. This is illustrated in 
\citefig{fig:comp_prop_mod}, where the solid red (gray) line shows the consequence of promoting the 
$\lambda \overset{E\to E_s}{\longrightarrow} 0$ regime as valid beyond the vertical boundary. The 
difference with the bounded case (dotted line) is hardly visible in the plot, and amounts only to 
a few percents.

We will adopt the so-called {\em med} set of propagation parameters 
\cite{2001ApJ...555..585M,2004PhRvD..69f3501D}. These parameters are supported by more 
recent constraints (\eg~\cite{2010A&A...516A..66P}), and could even be regarded as conservative
given complementary analyses which favor larger halo models 
(\eg~\cite{2011A&A...534A..54S,2012JCAP...01..049B}). We note that small halo models are likely
already excluded since they induce secondary positrons in excess with respect to the 
current data \cite{2011MNRAS.414..985L}.

\subsection{Ascribing energy Losses}
\label{subsec:losses}
Here, we summarize the way we compute the IC and synchrotron energy losses.
At high energies, the dominant component is gamma-ray emission through inverse 
Compton scattering. The target photons can be any of the three populations (CMB,
IR, SL) introduced in \citesec{sec:astro}. For each population, $i=$CMB, IR, SL, we calculate 
the energy loss rate per electron from a full relativistic treatment
\cite{1965PhRv..137.1306J,1968PhRv..167.1159J,1970RvMP...42..237B}:
\ben
\label{eq:bic}
b_i(E, r)&=& 3 \sigma_T \int_0^{\infty} {\rm d}E_{\rm in} \, 
E_{\rm in} \int_{1/4\gamma^2}^1 {\rm d} q \,
{{\rm d} n_i \over {\rm d} E_{\rm in}}(E_{\rm in}(q),r)
{(4\gamma^2-\Gamma_{e})q-1\over (1+\Gamma_{e} q)^3} \times\\
&&\left\{2q\log q +q+1-2q^2+{1\over 2}{(\Gamma_{e}q)^2\over 1+\Gamma_{e}q}(1-q)\right\} \, .
\nn
\een
Compared to the emitted power in \citeeq{eq:power}, one has to perform an additional integral over 
the energy of the outgoing photon, $E_{\gamma}$. In the Thomson regime where the incoming photon 
energy is much smaller than the electron mass in the electron rest frame (or, in the lab frame, 
$4E_{\rm in} E \ll m_e^2$), this formula simplifies substantially.

Electrons can also lose a substantial fraction of their energy through synchrotron 
radiation. The energy loss rate will crucially depend on the magnetic field in the Galaxy 
though. It is given by
\ben
\label{eq:bsync}
b_{\rm synch}(E, r)= {4\over 3} {\sigma_T E^2\over m_e^2} u_B(r) \, ,
\een
where the magnetic energy density was given in \citeeq{eq:ub}.

We show in \citefig{fig:EnLoss} both the energy dependence and the location dependence of the 
energy losses experienced by electrons in our MW.
\begin{figure}
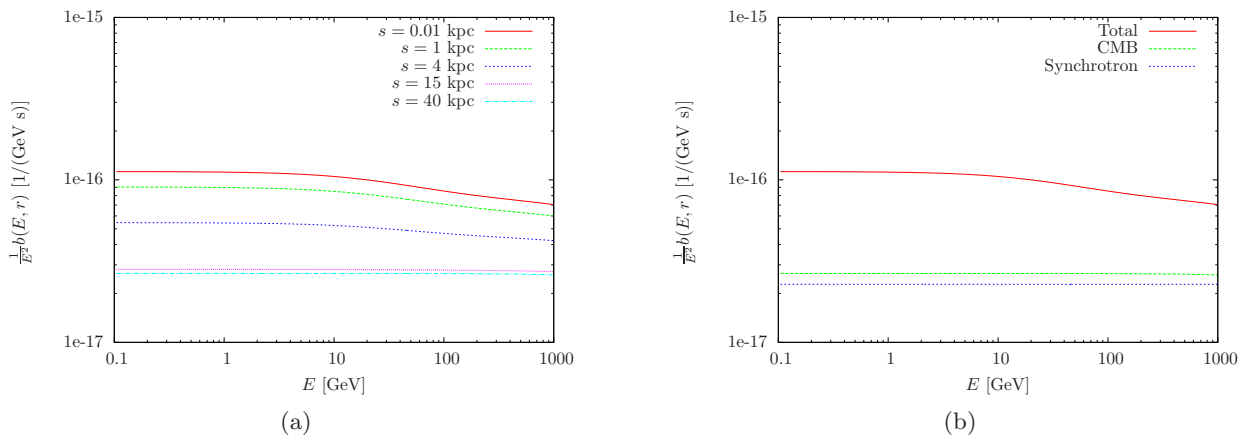

\hspace{-1cm}
\subfloat[]{\scalebox{0.6}{\input{Figures/EnLossbis_1000.tex}}}
\hspace{1cm}
\subfloat[]{\scalebox{0.6}{\input{Figures/EnLossCompbis_1000.tex}}}
\caption{Energy losses for an electron of energy $E$ at different positions on the 
l.o.s. $s$ towards the GAC (a), and at location $r=r_{\odot}$ for the different components (b).
In panel (b), the curve dubbed {\em Total} includes the other ISRF components 
(IR and SL) in addition to the CMB and the magnetic field --- see the text for more details.} 
\label{fig:EnLoss}
\end{figure}

Other losses such as Bremsstrahlung and ionization on interstellar matter are subdominant in the 
energy range of interest to us, and therefore we will not consider them here.

\section{DM Subhalos}
\label{sec:sub}
Structure formation is expected to occur hierarchically when the universe becomes matter-dominated.
For WIMP-like CDM particles, for instance most of supersymmetric models, typical free-streaming 
scales are such that very small structures can form, which correspond to (sub)halo masses down to 
$10^{-10}$-$10^{-4}$ $\msun$ (see \eg~\cite{2003PhRvD..68j3003B,2009NJPh...11j5027B}).

Since formed and virialized earlier than the bigger galactic halos, in a denser universe, these tiny
subhalos must be more concentrated --- this is mostly a qualitative picture, since in practice, 
structures from the smallest to intermediate scales form almost simultaneously. Anyway, these 
clumps should be numerous in their host galaxies, because they were already present at their 
later formation epoch or were accreted subsequently. Actually, these subhalos have long 
been observed on galactic scales in cosmological simulations, the most resolved of which being now 
able to characterize objects with masses down to $\sim 10^4\, \msun$ at redshift 0, with an 
impressive statistics amounting to hundreds of thousands of them (see the Via Lactea II 
\cite{2008Natur.454..735D} and Aquarius \cite{2008MNRAS.391.1685S} suites).

The presence of a large number of subhalos has important consequences in the way predictions
associated with the DM annihilation signals have to be derived. Indeed, because the annihilation 
rate scales with the squared DM density, the presence of inhomogeneities should increase the 
canonical predictions by a factor ${\cal B} \sim \langle n_\chi^2 \rangle / \langle n_\chi \rangle^2 >
1 $, where $n_\chi$ is the DM particle number density, and the average is performed
over the volume relevant to the specific detection channel \cite{1993ApJ...411..439S}.
This has long been recognized for the prompt emission, for which a boost factor with respect to 
canonical predictions was predicted, exhibiting a dependence on the angle with respect to the 
Galactic center (see \eg~\cite{1999PhRvD..59d3506B,2003PhRvD..68j3003B,2006NuPhB.741...83B,2011PhRvD..83b3518P}). 
In contrast, the contribution of inverse Compton processes relies on the distribution of the 
DM-induced electrons, which strongly departs from the DM distribution itself because of diffusion. 
To our knowledge, subhalo effects have never been included in this specific context in the past, 
while they might have important consequences for constraining leptophilic DM models\footnotemark. 
In the following, we present the method we have adopted to include subhalos in our calculation. It 
is actually based on previous studies on the DM-induced antimatter signals 
\cite{2007A&A...462..827L,2008A&A...479..427L,2011PhRvD..83b3518P}, to which we refer the reader 
for more details. Indeed, we first need to calculate the electron distribution in the whole 
Galactic halo.

\footnotetext{A similar implementation of subhalos can still be found in 
Ref. \cite{2010arXiv1008.1801Z}, though in the context of anisotropy studies.}

Assuming a universal DM density profile in all subhalos, the main statistical properties of 
the subhalo population are the mass function, the spatial distribution, and the concentration 
function. Here, we define the latter as $c \equiv r_{200}/r_{-2}$, where the $r_{200}$ 
is the radius at which the spherically averaged subhalo density is 200 times the 
critical density, and $r_{-2}$ the radius at which the logarithmic slope of the density profile
equals -2. The local number density can then be expressed as
\ben
\label{dNdVdM}
{d \Nsub (\vec{x}, \Msub)\over d V \, d \Msub} &=& \Nsubtot \,
{ d \prob{M}(\vec{x},\Msub) \over d \Msub} \,
{ d \prob{V} (\vec{x}) \over d V} \,
{ d \prob{c} (c,\vec{x},\Msub) \over d c} \\
&\simeq & \Nsubtot \,
{ d \prob{M} (\Msub) \over d \Msub} \,
{ d \prob{V} (r) \over d V} \, ,\nn
\een
where $\Nsubtot$ is the total number of subhalos in the Galaxy, and functions $\prob{}$ are
the probability density functions (pdfs). In the second line above, we have assumed (i) spherical
symmetry, (ii) that the mass distribution is spatially homogeneous, and (iii) that the concentration
parameter is fully fixed by the subhalo position and mass. The second approximation is not accurate
in the very center of the Galaxy where tidal effects disrupt massive objects, a region we are not
interested in in the present study, but still leads to a negligible effect in terms of global 
subhalo mass fraction. The third approximation cannot introduce spurious effects in the calculation 
of mean values, because the concentration pdf is usually found to be Gaussian for a given mass and 
radial position. The pdfs are all normalized to unity:
\ben
&& \int_{M_{\rm min}}^{M_{\rm max}} 
d \Msub \, { d \prob{M} (\Msub) \over d \Msub}   =1 \, ,\\
&& \int_{\rm MW} dV \, { d \prob{V} (r) \over  d V}  = 
4\, \pi \int_{0}^{R_{\rm vir}}dr\,r^2 \, { d \prob{V} (r) \over  d V} = 
1 \, .\nn
\een
We therefore need to ascribe a spatial distribution and a mass function to the subhalo population.
Following \cite{2011PhRvD..83b3518P}, we write the total mass density profile introduced in 
\citeeq{eq:densprof}) as the sum of a smooth component (`sm') and a clumpy one (`sub'),
\ben
\label{eq:smsub}
\rho_{\rm tot} (r) = \rhosm(r) + \rhosub(r) \, .
\een
Indeed, despite their limited resolution, it is reasonable to expect that the global density 
profiles found in current cosmological simulations, on galactic scales, will not vary in arbitrarily
finer ones, though $\rhosub$ would be then more constrained. Assuming that subhalos carry
a total MW mass fraction of $\fsub$, that we will discuss later, integrating the 
previous equation leads to
\ben
\label{eq:mass_contrib}
& 4\pi \int_0^{R_{\rm vir}} d r \, r^2 \, \rho_{\rm tot}(r) = \MMW  \,,\\
& 4\pi \int_0^{R_{\rm vir}} d r \, r^2 \, \rhosm(r) = (1-\fsub)\, \MMW  \,, \nn \\
& 4\pi \int_0^{R_{\rm vir}} d r \, r^2 \, \rhosub(r) = \fsub \, \MMW  \,.\nn
\een
If one further assumes that the smooth and subhalo components are radially antibiased, \ie~ 
$\rhosub \propto r \, \rhosm$ as found in \cite{2007ApJ...667..859D}, then consistency with 
\citeeq{eq:smsub} implies that \cite{2011PhRvD..83b3518P}
\ben
\rhosm (r)= {\rho_{\rm tot}(r) \over 1+r/r_b} \, ,\hspace{1cm}  
\rho_{\rm sub}(r)=  {\rho_{\rm tot}(r) \over 1+r/r_b}{r \over r_b} \, ,
\label{eq:biasrad}
\een
where $r_b$ is the bias radius, which can easily be calculated by injecting \citeeq{eq:biasrad} 
into \citeeq{eq:mass_contrib}.

The subhalo spatial pdf introduced in \citeeq{dNdVdM} is simply related to the substructure mass 
density as
\ben
{ d \prob{V} (r) \over  d V} = {\rhosub(r) \over \fsub\,  \MMW} \, .
\een

As to the mass distribution, it is usually found to be a power law consistent with the 
conventional Press-Schechter (PS) theory for gravitational collapse~\cite{1974ApJ...187..425P}. It 
may be expressed as
\ben
{  d \prob{M} (\Msub) \over d \Msub} (\Msub) = 
K_m \left[ { \Msub \over \msun }\right]^{-\alpha_m} \, ,
\een
where $K_m$ allows the normalization to unity inside the whole subhalo mass range:
\ben
K_m = {1 \over \msun} {\alpha_m - 1 \over \left({M_{\rm min} \over M_{\odot}}\right)^{1-\alpha_m} - 
\left({M_{\rm max}\over M_{\odot}}\right)^{1-\alpha_m} }\, .
\een

Apart from tidal stripping effects, which are important close to the galactic center, the mass pdf 
does not depend on the location in the galaxy. The accurate determination of the logarithmic
slope $\alpha_m$ is crucial for indirect dark matter searches given the broad mass range under
scrutiny; values from 1.9 to 2 already induce big changes in terms of global subhalo
luminosity~\cite{2008A&A...479..427L}. The PS theory predicts $\alpha_m=2$, which 
is confirmed in more general approaches (\eg~\cite{2001MNRAS.323....1S}) and close to what is 
found in cosmological simulations, the most resolved of which 
provide values spanning the range 1.9-2 (\eg~\cite{2008Natur.454..735D,2008MNRAS.391.1685S}).

We provide some typical values of the total subhalo number, mass fraction, \etc, (see discussion 
above), in \citetab{tab:table2}.

\begin{table}
\centering
  \begin{tabular}{l | c | c }
   $\alpha_m$ & $M_{\rm min}=10^{-11}M_{\odot}$ & $M_{\rm min}=10^{-4}M_{\odot}$ \\ \hline
     & $f^{\rm tot}_{\rm sub}= 0.699$  & $f^{\rm tot}_{\rm sub}= 0.467$ \\
   2 & $N^{\rm tot}_{\rm sub}= 2.66\times 10^{21}$  & $N^{\rm tot}_{\rm sub}= 2.66\times 10^{14}$ \\
   \hline
    & $f^{\rm tot}_{\rm sub}= 0.187$ &  $f^{\rm tot}_{\rm sub}= 0.181$ \\
    1.9 & $N^{\rm tot}_{\rm sub}= 3.06\times 10^{19}$ & $N^{\rm tot}_{\rm sub}= 1.54\times 10^{13}$\\ 
    \hline
  \end{tabular}
\caption{Subhalo parameters --- see the text for details.}
\label{tab:table2}
\end{table}

\subsection{Prompt emission}
For the prompt emission, {\em unresolved} subhalos are accounted for such that the l.o.s. integral 
of \citeeq{eq:prompt} is modified as follows (\eg~\cite{1999PhRvD..59d3506B,2006NuPhB.741...83B,2008MNRAS.384.1627P,2010PhRvD..81d3532K,2011PhRvD..83b3518P}):
\ben
\label{eq:prompt_sub}
{ {\rm d}\Phi \over {\rm d} E_{\gamma}} = { r_{\odot} \over 4\pi} \, {\cal S} \, 
{{\rm d} N_{\gamma}\over {\rm d} E_{\gamma}} \int {\rm d}\Omega \int_{\rm los} {{\rm d}s\over r_{\odot}}
\, {{\rm d} {\cal P}_V  \over {\rm d} V}(r) \int_{M_{\rm min}}^{M_{\rm max}(s)} 
{\rm d} \Msub \, \xi(\Msub, r) \, { {\rm d} \mathcal{P}_M (\Msub) \over {\rm d} \Msub}\, ,
\een
where the {\em local} effective annihilation volume is defined as \cite{2008A&A...479..427L}
\ben
\xi(\Msub, \vec{x}_s)\equiv \int_{V_{\rm sub}} {\rm d} V 
\left({\rho_{\rm cl}(\Msub, \vec{x}_s) \over \rho_{\odot} }\right)^2 \, ,
\een
which depends on the density profile $\rho_{\rm cl}$ of the clump centered at position $ \vec{x}_s$ 
and extending over the volume $V_{\rm sub}$. This corresponds to the volume that would provide
the same global annihilation rate if the DM density inside the subhalo was taken constant and
fixed to the solar value.

For an NFW profile, the annihilation volume has an analytic expression in terms of the 
concentration parameter $c$ \cite{2008A&A...479..427L}:
\ben
\xi(\Msub, \vec{x}_s) = {\Msub^2 \over 12\pi \rho_{\odot}^2 r_{200}^3} \, 
{c^4(3+c(3+c)) \over
(1+c)(c-(1+c)\log(1+c))^2}\,.
\een
We note that a complementary approach to determine the subhalo contribution to the prompt 
gamma-ray emission is available in the form of a public Monte Carlo code, called CLUMPY 
\cite{2012CoPhC.183..656C}. Such a code should in principle reproduce our results for the
prompt emission (provided the same input parameters are used), but does not contain any CR 
transport module, without which no IC calculation can be performed.
\subsection{Inverse Compton}
The inverse Compton component depends crucially on the electron population coming from DM 
annihilation. In addition to the prompt component for which a mere line-of-sight integral is 
performed, here one should also include a volume integration to take into account the diffusion of 
these electrons and determine their density at each line-of-sight step. The electron density 
originating in Galactic DM clumps is then given by
\ben
{\cal N}^{\rm sub}_e( E,\vec{x}) &=&{\cal S}  \int {\rm d}^3 \vec{x}_s 
\widetilde{\cal G}_e(E,\vec{x}\leftarrow \vec{x}_s)\, 
{{\rm d} \mathcal{P}_V  \over {\rm d} V}(\vec{x}_s) \\
&& \times \int_{M_{\rm min}}^{M_{\rm max}} 
{\rm d} \Msub \, \xi(\Msub, \vec{x}_s) \, { {\rm d} {\cal P}_M (\Msub) 
\over {\rm d} \Msub}  \, ,\nn\\
{\rm with} && \widetilde{\cal G}_e(E,\vec{x}\leftarrow \vec{x}_s) \equiv \int_E^\infty dE_s \,  
{\cal G}_e(E,\vec{x}\leftarrow E_s, \vec{x}_s)\,   {dN_e(E_s) \over dE_s}\,,\nn
\een
which has to be substituted in \citeeq{eq:fluxIC} in order to find the gamma-ray flux coming from 
the Galactic subhalo population.

\section{Results}
\label{sec:res}
In the previous sections, we introduced all the ingredients necessary to compute the gamma-ray flux 
on Earth from DM annihilation in the Galaxy. We showed how the different components make up the 
total flux. Annihilation can originate from the smooth Galactic DM halo, or from Galactic 
subhalos. Moreover, photons can be produced either promptly, or via IC scattering, notably when 
light leptons make a significant part of the annihilation products.

In \citefig{fig:Fluxtau} we put everything together and show the total diffuse gamma-ray flux coming
from a 1 TeV DM particle annihilating into $e^+e^-$ (top) and $\tau^+\tau^-$ (bottom) within our 
Galaxy for two directions in the sky: the galactic anticenter (GAC) and the Galactic pole(s) 
(GP). The GAC is given by Galactic coordinates $(b,\ell)=(0,180^{\circ})$, and it corresponds to the 
direction where the flux is expected to be the lowest. In that respect, it can be thought of as the 
irreducible flux for all directions. GP is defined as the direction $(90^{\circ},0)$, and we choose 
it because the astrophysical Galactic diffuse emission (the main foreground) is best predicted
and controlled in this direction. We have fixed the annihilation cross-section to the typical 
thermal relic one\footnotemark, $\sigv=3\times 10^{-26}$~cm$^3$/s, and contrasted this flux with the 
observed IGRB obtained by Fermi-LAT with the first 10 months of data~\cite{2010PhRvL.104j1101A}. 
These first data extend up to 100~GeV, and we have added new preliminary data which extend the 
measurement of the IGRB up to 600~GeV~ in the Figure~\cite{Morselli}. Note that the determination 
of the IGRB relies on the modeling of the diffuse emission within our Galaxy, and the identification
of sources by Fermi-LAT, which are subtracted. For the smooth halo (IC only) contribution, we 
compare two calculation results: one which neglects the diffusion of electrons (dotted lines), and 
another one which properly includes it (solid lines).

\footnotetext{The notion of {\em typical} has to be taken with caution, since it is well-known
that light WIMPs should be associated with about twice larger annihilation cross sections than heavy
WIMPs, with a transition occuring when the thermal decoupling temperature is close to
that of the QCD phase transition, where the number of relativistic degrees of freedom---and 
thereby the expansion rate of the universe---experiences
a rapid change~\cite{2012NuPhB.854..738C,2012PhRvD..86b3506S}.}

Looking at the upper panels of \citefig{fig:Fluxtau}, associated with the annihilation into 
$e^+e^-$, we first observe that the three IC bumps are salient, corresponding, from left to right, 
to the electrons scattering off CMB photons, IR light, and SL. The dominant bump is clearly seen to 
be the CMB one. We also notice that the relative contribution from IC scattering is not negligible 
compared to the prompt component, although the prompt emission will still dominate the exclusion 
limits on the annihilation cross-section we will derive below. On the other hand, for 
$\tau^+\tau^-$ (lower panels of \citefig{fig:Fluxtau}), we see that the prompt 
component is much more prominent than IC, and it will be clearly dominant when setting bounds on 
the annihilation cross-section. As discussed in \citesec{sec:gamma}, the number of photons emitted 
promptly in this mode of annihilation is larger because of the hadronic component in $\tau$ decays.

Second, it is clear, in particular from the left panels of \citefig{fig:Fluxtau} (GP),
that including spatial diffusion is very important for proper predictions of the IC contribution.
Neglecting diffusion leads to an overestimate of the low energy GP flux by as much as a factor of 
2, where the CMB contribution is dominant. This is due to the dilution of the electron density 
beyond the half-height of the diffusion slab (4~kpc). On the other hand, for the GAC the effect of 
diffusion is barely noticeable  as the radius of the diffusion zone is large (20~kpc), and most of 
the gamma-ray signal is contained within that distance.

Finally, in order to illustrate the effect of unresolved subhalos, we show in \citefig{fig:Fluxtau} 
the differential gamma-ray flux after integration along the line of sight, fixing $\alpha_m=2 $ and 
$M_{\rm min}=10^{-6}M_{\odot}$, and we find that the flux can be enhanced by up to a factor 3 for the 
GP and 5 for the GAC.
\begin{figure}[t]
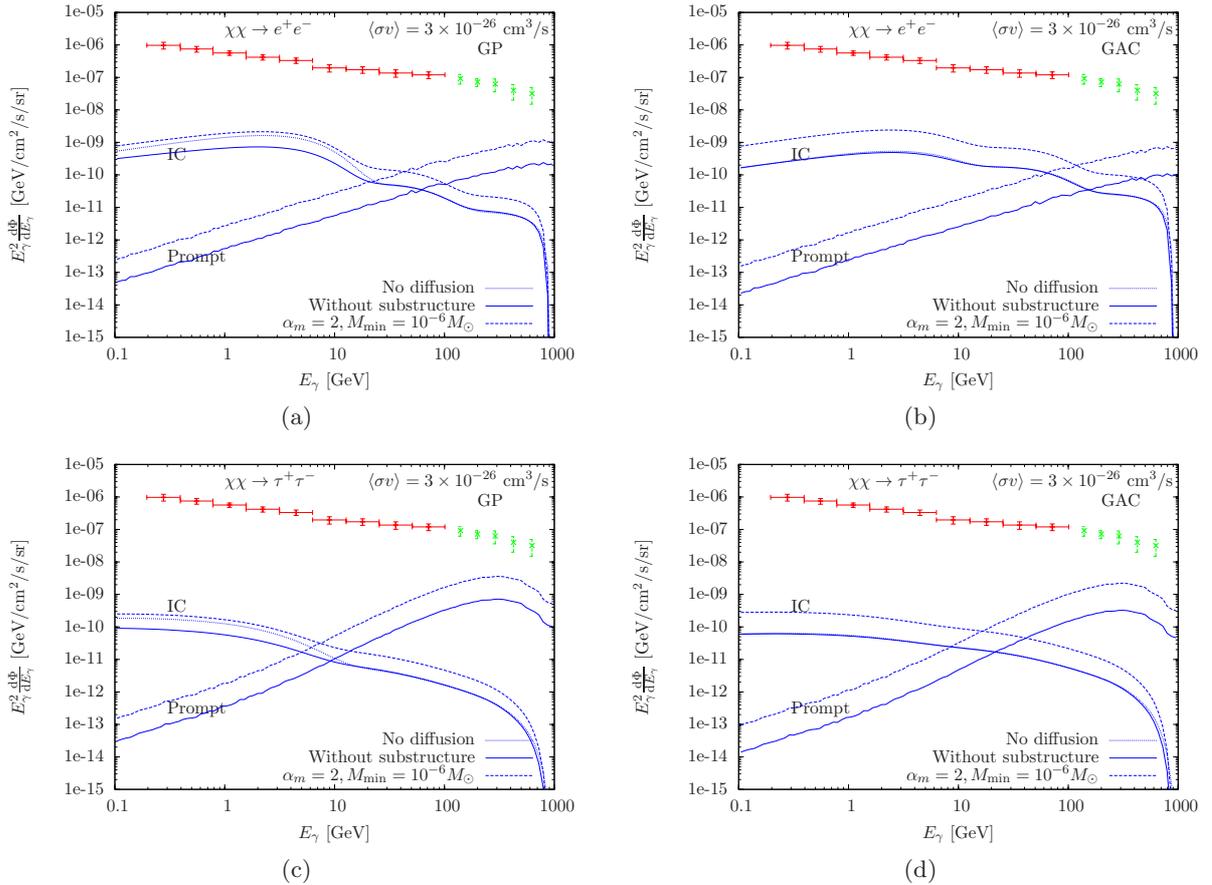

  \subfloat[]{\scalebox{0.6}{\input{Figures/FluxHLLee989.tex}}}
\hspace{0.5cm}
\subfloat[]{\scalebox{0.6}{\input{Figures/FluxAGCee989.tex}}}
\vspace{0.1cm}
\subfloat[]{\scalebox{0.6}{\input{Figures/FluxHLLtau989.tex}}}
\hspace{0.5cm}
\subfloat[]{\scalebox{0.6}{\input{Figures/FluxAGCtau989.tex}}}
  \caption{Differential gamma-ray flux for a 1 TeV DM particle annihilating into $e^+e^-$ (top) and
    $\tau^+\tau^-$ (bottom) in the direction of the GP (left) and GAC (right). Shown are the IC and 
    the prompt components, with and without boost due to subhalos.} 
  \label{fig:Fluxtau}
\end{figure} 

In \citefig{fig:FluxVare2} we further illustrate the dependence of the gamma-ray flux 
on the main subhalo parameters $\alpha_m$ and $M_{\rm min}$ for the DM annihilation channels 
$e^+e^-$ and $\tau^+\tau^-$. We vary $M_{\rm min}$  between $10^{-11}M_{\odot}$ and $10^{-4}M_{\odot}$, 
which can be motivated from the point of view of the kinematic decoupling of WIMPs (see 
\cite{2009NJPh...11j5027B} and references therein). Regarding the mass function index $\alpha_m$ 
we take as reference values $\alpha_m=2$ and $\alpha_m=1.9$, which are motivated from spherical 
collapse models and N-body simulations, respectively, as discussed in \citesec{sec:sub}. Although 
these indices seem rather close, the implications for the gamma-ray flux are dramatic. With 
$\alpha_m=1.9$, the signal is very marginally enhanced by the presence of subhalos, and there is a 
very mild dependence on the choice of minimal subhalo mass. On the other hand, for $\alpha_m=2$, 
the minimal subhalo mass is crucial and the expected signal can vary by up to a factor 5. This was 
already emphasized in previous studies 
(\eg~\cite{2008A&A...479..427L,2008MNRAS.384.1627P,2011PhRvD..83b3518P,2012CoPhC.183..656C}).
\begin{figure}[t]
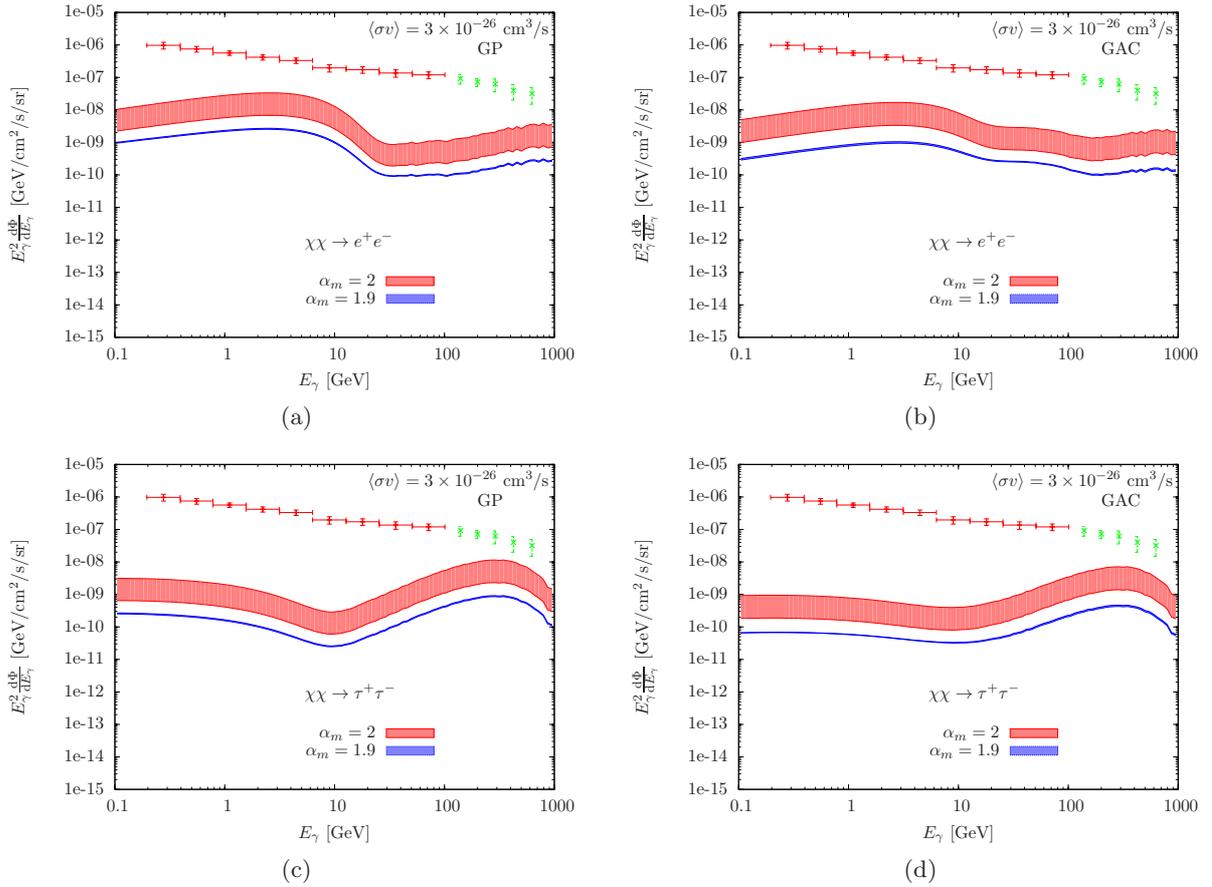

  \subfloat[]{\scalebox{0.6}{\input{Figures/FluxRangeHLL.tex}}}
\hspace{0.5cm}
\subfloat[]{\scalebox{0.6}{\input{Figures/FluxRangeAGC.tex}}}
\vspace{0.1cm}
  \subfloat[]{\scalebox{0.6}{\input{Figures/FluxRangeHLLtau.tex}}}
\hspace{0.5cm}
\subfloat[]{\scalebox{0.6}{\input{Figures/FluxRangeAGCtau.tex}}}
  \caption{Differential gamma-ray flux for a 1 TeV DM particle annihilating into $e^+e^-$ (top)
    and $\tau^+\tau^-$ (bottom) including the contribution from subhalos
    in the direction of the GP (a)--(c) and GAC (b)--(d). The colored bands correspond
    to varying the minimal subhalo mass between 
    $10^{-11}$ and $10^{-4} M_{\odot}$ .} 
  \label{fig:FluxVare2}
\end{figure} 

It is interesting to fully extract the signal enhancement coming from DM subhalos. Here we define 
the effective {\em boost factor} as
\ben
{\rm Boost}\equiv { {\rm d}\Phi^{\rm sub}/{\rm d} E_{\gamma}+{\rm d}\Phi^{\rm smooth}/{\rm d} E_{\gamma} 
\over {\rm d}\Phi^{\rm no sub}/{\rm d} E_{\gamma}} \,,
\een
where $\Phi^{\rm sub}$ is the flux coming from subhalos (both through prompt and IC emissions), 
and $\Phi^{\rm smooth}$ is the flux coming from the smooth part of the DM halo. Note that the 
splitting between smooth halo and subhalos is done in a self-consistent way: as explained in detail 
in \citesec{sec:sub}, we keep the total MW mass constant and the density profile at any radius 
satisfies \citeeq{eq:densprof}. The result is shown in \citefig{fig:Boost} for the GP and the
GAC. We vary the substructure parameters as in the previous figure. When $\alpha_m=1.9$, we 
find that the signal is only enhanced by 20-30$\%$ and 30-40$\%$ for the GP and the GAC, 
respectively, with very little dependence on $M_{\rm min}$. On the other hand, when $\alpha_m=2$, we 
find that the boost can be up to a factor 20, the highest value being obtained for the prompt 
emission towards the GAC and for the lowest minimal subhalo mass $M_{\rm min}= 10^{-11} M_{\odot}$. 
We also notice that the boost factor is larger for the prompt component than for IC, because the 
dilution effect beyond the diffusion slab mentioned above implies that only nearby substructures 
have an impact on the local IC gamma-ray flux --- CR electrons are strongly diluted outside the 
diffusion zone. Note that this effect is especially strong in the GP direction, because of the 
small half-height compared to the radius of the diffusion zone. We will discuss the issue of 
transport outside the diffusion zone into more details in a forthcoming paper, where the impact of 
the anisotropic electron flux will be evaluated.

The boost has a non-trivial energy dependence which is apparent for the IC component: We can 
clearly see the different IC bumps as in the figures depicting the gamma-ray flux. The 
largest boost is always linked to the CMB bump, because this is where the diffusion length is the 
largest, and therefore the integration volume includes more substructures. On the other hand, the 
boost for the prompt signal is, as already well established \cite{1999ApJ...526..215B}, independent 
of energy. We also notice by comparing panels (a) and (b) of \citefig{fig:Boost} that the boost 
factor is always larger in the case of the GAC compared to GP. The reason is that the relative 
mass domination of the subhalo component is reached faster towards the GAC  (the line-of-sight 
variable $s= r-r_\odot$), than towards the GP ($s = \sqrt{r^2 - r_\odot^2}$), and therefore the boost 
is correspondingly larger. This is important as it implies that unresolved substructures tend to 
render the global signal more isotropic than a purely smooth one, as noticed in 
\cite{2010JCAP...04..014A} for instance. In our case, the signal from the GAC is the lowest of all 
directions, but it is partially compensated by a larger boost. As it can be noticed in 
\citefig{fig:FluxVare2} for the band $\alpha_m=2$, the compensation is not complete but the effect 
is clearly visible.
\begin{figure}[t]
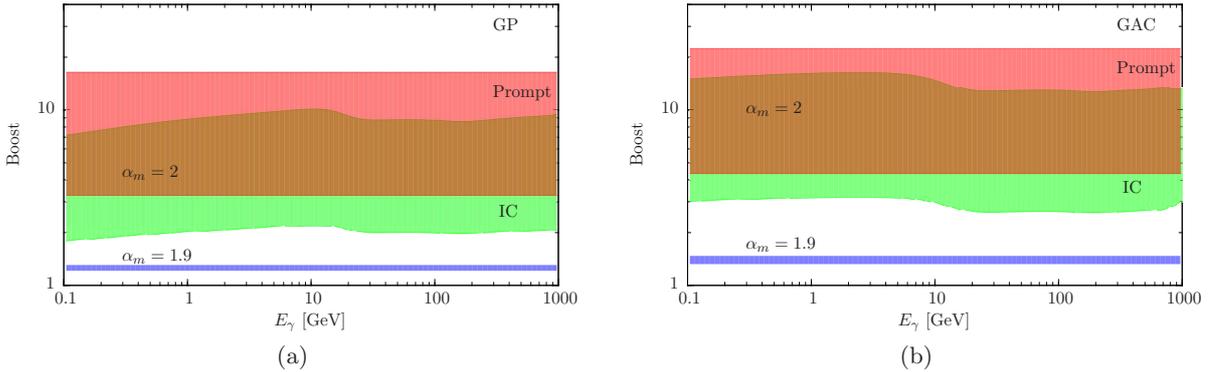

\subfloat[]{\scalebox{0.6}{\input{Figures/BoostLogHLLDiffMED.tex}}}
\hspace{0.5cm}
\subfloat[]{\scalebox{0.6}{\input{Figures/BoostLogAGCDiffMED.tex}}}
  \caption{Boost factor as a function of the energy for a 1 TeV DM particle annihilating into 
    $e^+e^-$, in the direction of GP (a) and GAC (b). 
    The upper colored regions describe the range of boosts for the prompt (red/dark shaded) 
    and IC components (green/light-shaded) varying the minimal halo mass between $10^{-11}$ and 
    $10^{-4}\,\msun$ for $\alpha_m=2$. The lower flat region (or thick line) represents the range 
    of boosts for the prompt component with $\alpha_m=1.9$ for the same interval of minimal halo 
    masses.} 
  \label{fig:Boost}
\end{figure} 

Now that we have calculated the total gamma-ray flux from the prompt and IC components, 
we can 
establish exclusion limits on the DM annihilation cross section as a function of the WIMP mass, 
for different annihilation channels.  We use the IGRB as our benchmark data including the 
preliminary data points at higher energies shown in \citefig{fig:Fluxtau}~\cite{Morselli}. Our 
bounds will be set at 2$\sigma$ following the statistical procedure outlined 
in~\cite{2010JCAP...03..014P}, where the $\chi^2$ is conservatively calculated including only
the energy bins where the signal is larger than the IGRB measurement. We will later see how these 
bounds get stronger when the DM signal is combined with an astrophysical explanation for the IGRB 
in the form of a power law. 

\begin{figure}[p!]
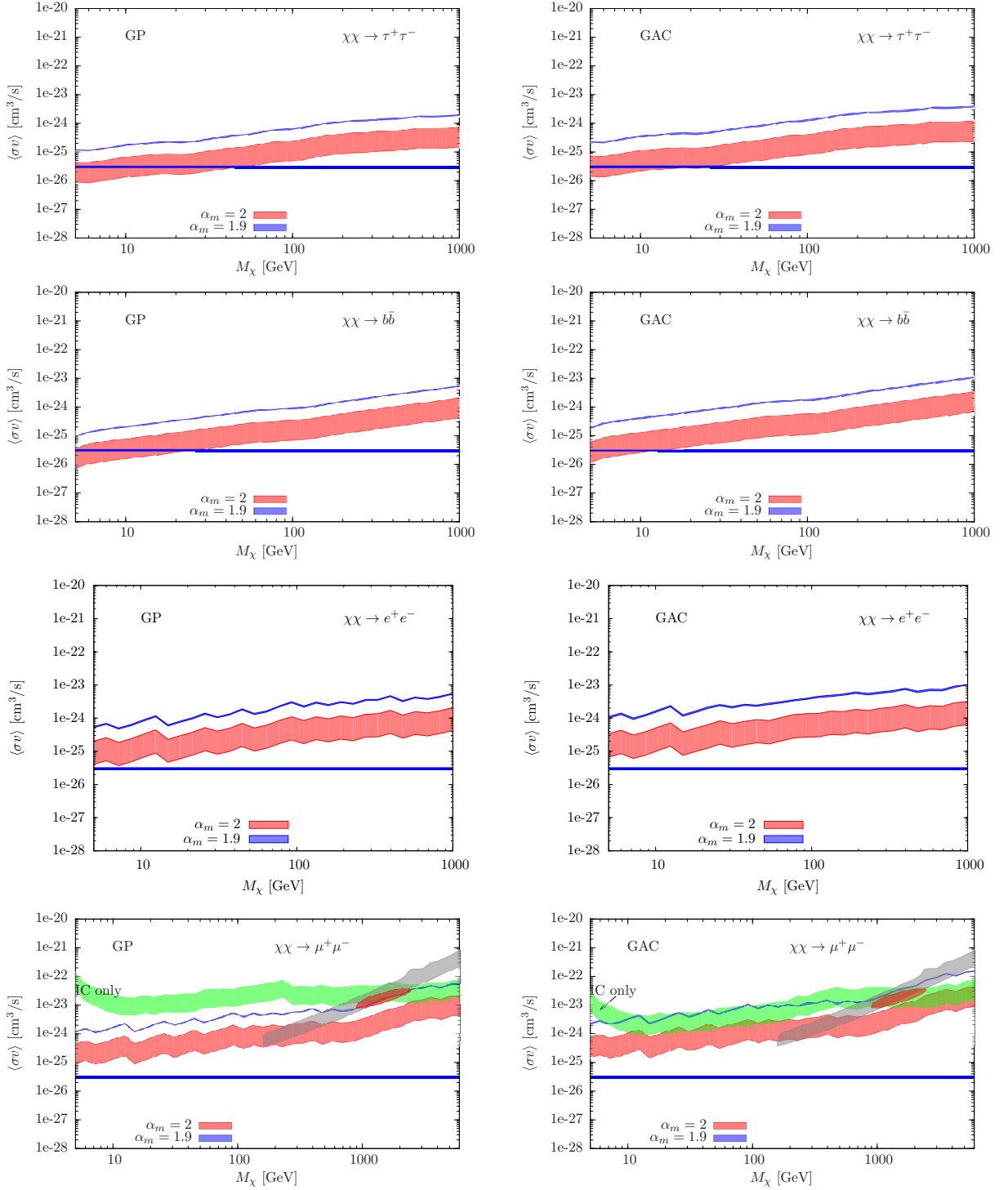

\vspace{-1cm}
 {\scalebox{0.6}{\input{Figures/LimitHLLtauNew.tex}}}
\hspace{0.5cm}
 {\scalebox{0.6}{\input{Figures/LimitAGCtauNew.tex}}}
\vspace{0.1cm}
{\scalebox{0.6}{\input{Figures/LimitHLLbbNew.tex}}}
\hspace{0.5cm}
{\scalebox{0.6}{\input{Figures/LimitAGCbbNew.tex}}}
\vspace{0.1cm}
{\scalebox{0.6}{\input{Figures/LimitHLLee.tex}}}
\hspace{0.5cm}
{\scalebox{0.6}{\input{Figures/LimitAGCee.tex}}}
\vspace{0.1cm}
{\scalebox{0.6}{\input{Figures/LimitHLLmm.tex}}}
\hspace{0.5cm}
{\scalebox{0.6}{\input{Figures/LimitAGCmm.tex}}}
  \caption{Limits on the thermally averaged DM annihilation cross-section for different channels,
in the direction of GP (left) and GAC (right) when $\alpha_m=2$ and $\alpha_m=1.9$. In the bottom
panels, the orange and gray ellipses represent the regions taken from ~\cite{2010NuPhB.831..178M} 
favored to explain the positron excess (at 3$\sigma$) by PAMELA~\cite{2009Natur.458..607A} and the
excess in $e^++e-$ seen by Fermi-LAT ($5\sigma$)~\cite{2009PhRvL.102r1101A}, respectively.} 
  \label{fig:Limitstau}
\end{figure}  

It does not come as a surprise that the strongest limits are obtained for annihilation into 
$\tau^+\tau^-$, as shown in the upper two plots of \citefig{fig:Limitstau}. 
For the GP direction, we notice that the typical cross section for a thermal relic with 
$\alpha_m=2$ and $M_{\rm min}\lesssim 10^{-6}M_{\odot}$ is disfavored in this channel up to DM masses 
of about 30~GeV. On the other hand, if the index is $\alpha_m=1.9$, the thermal relic 
cross section is not excluded for any mass. 

The channel $b\bar{b}$ is displayed under the $\tau^+\tau^-$ case in \citefig{fig:Limitstau}, 
and one can see that it is also partially constrained down to the thermal relic cross section at 
low DM masses when $\alpha_m=2$. On the other hand, the leptonic channels $e^+e^-$, $\mu^+\mu^-$
never reach the thermal relic cross-section, even for $\alpha_m=2$. However, in the case 
$\mu^+\mu^-$, it can be seen that the regions favored to explain the 
PAMELA~\cite{2009Natur.458..607A} and Fermi-LAT~\cite{2009PhRvL.102r1101A} excesses are marginally 
allowed when $\alpha_m=1.9$ and disfavored when $\alpha_m=2$. Note that we rescaled the regions 
found in \cite{2010NuPhB.831..178M} in order to account for a different local DM density 
$\rho_{\odot}=0.395$~GeV/cm$^3$ vs.  0.3~GeV/cm$^3$. On the other hand, we did not account for a 
potential boost factor of the electron-positron flux due to substructures, which would move the 
regions by a factor 2-3 downwards in the case $\alpha_m=2$ (no change when 
$\alpha_m=1.9$)~\cite{2011PhRvD..83b3518P}.

We now comment on the relative importance of the prompt vs. IC emission for the limits. It is clear 
that the hadronic channels $\tau^+\tau^-$ and $b\bar{b}$ are heavily dominated by the prompt 
emission. But also in the purely leptonic cases the prompt component dominates the signal and 
therefore the bounds all the way up to 600 GeV, leaving little room for IC. 
In previous 
works~\cite{2010JCAP...03..014P,2010NuPhB.840..284C,2010JCAP...07..008H,2012PhRvD..85d3509A}, the 
IC component was clearly dominating the exclusion limits above 500 GeV. This discrepancy is easily 
explained by the choice of data sets. With only data published by Fermi-LAT collaboration in 
\cite{2010PhRvL.104j1101A} we have indeed that the prompt component does not give a strong bound 
for DM masses above 500 GeV, because the flux peaks far beyond the energy range of this data set.
However, including the preliminary data up to 600 GeV implies that the prompt component is dominant 
up to higher energies.

A comparison with previous works is now in order. As we mentioned earlier, a direct comparison is 
often complicated because exclusion limits are often extracted with different assumptions: 
including galactic and/or extragalactic contribution, with different data sets or directions in 
the sky, and finally the exclusion limits sometimes refer to $1\sigma$, $2\sigma$ or $99\%$ C.L. 
In the study made by the Fermi collaboration~\cite{2010JCAP...04..014A} only the extragalactic 
signal from DM annihilation was used to derive limits. The data set was the 
IGRB~\cite{2010PhRvL.104j1101A}. Compared to the $95\%$ C.L. found in their work, our limits
for $\alpha_m=1.9$ are close to their reference conservative limit (MSII-Sub1). As we can see
in \citefig{fig:Limitstau}, our best limits in the case $\alpha_m=2$ are about one order of 
magnitude lower, which makes them slightly stronger than the case denoted by `BulSub' 
in~\cite{2010JCAP...04..014A}. Note however that the cosmological signal from DM annihilation is 
even more sensitive to the DM substructure distribution than the galactic signal, as it can vary 
by up to three orders of magnitude (see for instance Fig.~1 in~\cite{2010JCAP...04..014A}) 
compared to one order of magnitude in our case.

In~\cite{2012PhRvD..85d3509A}, the Galactic signal in the direction of the Galactic anticenter 
was considered, and compared to the IGRB. It agrees well with our result when galactic substructure
is neglected. In~\cite{2010JCAP...07..008H}, both the Galactic and extragalactic signals were 
computed and compared to the IGRB. Although no subhalo enhancement was included for the Galactic 
contribution, it was found to be typically dominant over the extragalactic one except for rather 
extreme concentration models. We obtain consistent results with this work when neglecting subhalo 
boost.

To date, the robust limits in the field of DM indirect detection with photons which are strongest
at low DM masses are those obtained from a combined analysis of the DM signal from
a set of Dwarf Spheroidal galaxies which are satellites of our 
MW~\cite{2011PhRvL.107x1302A,2011PhRvL.107x1303G}. In \cite{2011PhRvL.107x1303G}, 
it was found that the thermal relic cross section for a WIMP was starting to be excluded for low 
masses in the channels $\tau^+\tau^-$ and $b\bar{b}$. Their result is slightly better than 
our best limits in the direction of the GP with the most favorable choice of parameters 
($\alpha_m=2$ and $M_{\rm min}=10^{-11}M_{\odot}$). Note that our limits are more constraining at 
higher masses $M_{\chi}>100$~GeV because we are using the additional preliminary data points 
of~\cite{Morselli}.

The limits on DM annihilation cross sections shown in \citefig{fig:Limitstau} were obtained in
a conservative way. Although the shape of the DM signal is unlikely to resemble the power law
spectrum of the IGRB, we did not try to superimpose a model for the IGRB in order to 
derive bounds on DM annihilation cross-sections. Nevertheless, a few astrophysical sources were 
shown to be potentially at the origin of (at least a part of) the IGRB, most prominently 
blazars~\cite{2009ApJ...702..523I,2011ApJ...728...73I,2011PhRvD..84j3007A}, non-blazar Active 
Galactic Nuclei (AGN)~\cite{2008ApJ...672L...5I}, star-forming galaxies~\cite{2010ApJ...722L.199F}
and milli-second pulsars~\cite{2010JCAP...01..005F}. It is likely that a combination of 
some of these contributions explains the totality of the IGRB consistently with the anisotropy
constraints (\eg~\cite{2012PhRvD..86f3004C,2012arXiv1206.4734H}). Here we follow the strategy 
employed in \cite{2010JCAP...04..014A} to derive `stringent' limits. More specifically, we first
find the single power law which provides the best fit to the IGRB, which we assume to be a fixed
background. Then the `stringent' bounds are obtained by adding the DM contribution and extracting 
the annihilation cross section for which the $\chi^2$ deviation exceeds the classical 
2$\sigma$ limit. The result is shown in \citefig{fig:LimitstauBlazar} for 
the case of DM annihilating into  $\tau^+\tau^-$, and for the direction of the GP. The limits get a 
factor 5-6 stronger than in \citefig{fig:Limitstau}, with masses up to 100~GeV excluded if 
$\alpha_m=2$ and $M_{\rm min}=10^{-11}M_{\odot}$.
\begin{figure}[t]
  \begin{center}
    \scalebox{0.8}{\input{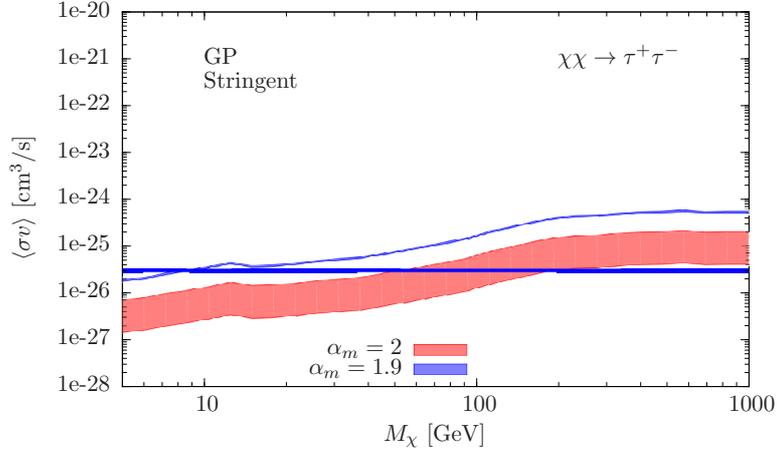}}
  \end{center}
  \caption{Stringent limits on DM annihilation cross-section for the channel $\tau^+\tau^-$ in the
    direction of GP obtained by superimposing a power law background to the DM signal (see text for
details).} 
  \label{fig:LimitstauBlazar}
\end{figure}  

We saw in the upper panels of \citefig{fig:Limitstau} that if DM annihilates mainly in channels 
$\tau^+\tau^-$ or $b\bar{b}$, {\em and at the canonical rate}, light masses are excluded 
when $\alpha_m=2$ and the minimal halo mass 
is $\lesssim 10^{-6}M_{\odot}$. This means that in these channels we have an interesting connection 
between the DM mass and the mass function index $\alpha_m$. We can thus exclude part of the 
parameter space $(M_{\chi},\alpha_m)$, as shown in \citefig{fig:alphavsmchi} for the case 
$\tau^+\tau^-$ in the direction of the GP. A substantial fraction of the parameter 
space is excluded at $2\sigma$, especially with the stringent limits as derived in 
\citefig{fig:LimitstauBlazar}, and when the minimal halo mass is as small as 
$10^{-11}M_{\odot}$. Therefore, if we came to know the DM mass from the Large Hadron Collider, and 
that it mainly annihilates into hadronic channels such
as $\tau^+\tau^-$ or $b\bar{b}$, we would learn something important about the mass function,
which is a central concept in the theory of structure formation. Conversely, if the mass function
for our galaxy is determined more accurately with N-body simulations, and the extrapolation
down to small halo masses is on a stronger footing,
we would obtain a robust exclusion of hadronic final states at low DM masses, depending on the 
value of the mass function index found.
\begin{figure}[t]
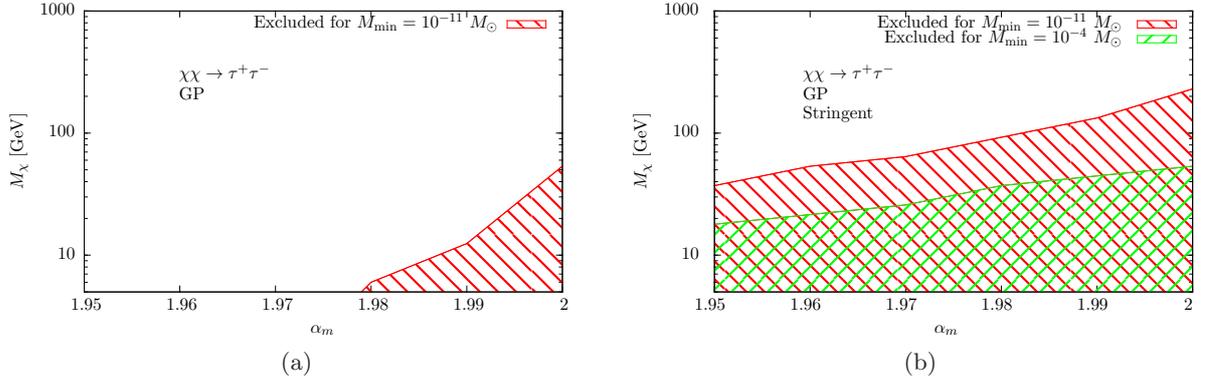

  \subfloat[]{\scalebox{0.6}{\input{Figures/AlphavsMchiNoBlazarNew.tex}}}
\hspace{0.5cm}
\subfloat[]{\scalebox{0.6}{\input{Figures/AlphavsMchiNew.tex}}}
  \caption{Exclusion regions at 2$\sigma$ in the plane $(M_{\chi},\alpha_m)$, if DM annihilates 
    {\em at the canonical rate} into
 $\tau^+\tau^-$. The left panel was obtained following the conservative procedure used in 
\citefig{fig:Limitstau}, whereas the right one was obtained following the `stringent' procedure
used in \citefig{fig:LimitstauBlazar}.
The minimal halo mass $M_{\rm min}$ is varied between $10^{-11}$ and $10^{-4}M_{\odot}$. } 
  \label{fig:alphavsmchi}
\end{figure}  

\section{Conclusions}
\label{sec:concl}
We discussed gamma-ray signals from DM annihilations in our Galaxy. We paid particular
attention to the following aspects which were neglected in previous analyses: 
(i) when DM annihilates into light leptons, the diffusion of the electrons/positrons before
IC scattering is potentially important, and (ii) substructures within our Galaxy can give a
significant additional contribution to the signal. We have first shown that diffusion effects
are more important in the calculation of the DM-induced diffuse gamma-ray flux towards high 
latitudes than towards the Galactic anticenter, due to the finite vertical extent of the 
confinement zone. This mostly impacts the low energy IC contribution, while the prompt emission,
for which CR transport is irrelevant, turns out to almost always be dominant.

Second, we have quantified the uncertainty in the gamma-ray signal due to the presence of 
subhalos in our Galaxy. In order to choose realistic parameters for the substructure
distribution, we used results from the latest N-body simulations, as well as some theoretical
arguments. This led us to choose a mass function index for our Galaxy between 1.9 and 2. 
Concerning the other important parameter that determines the substructure boost factor, namely the 
minimal halo mass, we considered a wide range of values which can be motivated by the kinetic
decoupling temperature of different DM particles. We found that under optimistic assumptions 
about both the minimal halo mass and the mass function index, the signal can be enhanced by a 
factor of 2-20, whereas with very conservative assumptions it could be as low as 20-30$\%$. 
Therefore, we have a theoretical uncertainty of roughly an order of magnitude 
in the prediction of the flux from DM annihilation. This can be found to be enormous, but it
is actually much less than the uncertainty in the prediction of the extragalactic signal, which 
can vary by three orders of magnitude.

In order to set exclusion limits on annihilation cross-sections, we used the Isotropic
Gamma-Ray Background measurement by Fermi-LAT. In the most optimistic subhalo scenario,
we obtain limits that are as stringent as those derived from Dwarf-Spheroidals
\cite{2010ApJ...712..147A,2011PhRvL.107x1303G,2011PhRvL.107x1302A}, with
the low mass region $M_{\chi}<30$~GeV excluded in the channel $\tau^+\tau^-$, 
assuming a canonical annihilation cross section.

Finally, we have shown that there exists an interesting connection between the mass function
index and the DM mass. In case of determination of the DM mass at a collider, and provided
the main annihilation channel is known to be hadronic, we will learn
something about the mass function index from the non-observation of DM signals in Fermi-LAT
data.

\begin{acknowledgments} 
  We would like to thank Oleg Ruchayskiy and Kevork Abazajian for useful discussions. SB 
  acknowledges support from the Swiss National Science Foundation, under the Ambizione grant 
  PZ00P2\_136947. JL wishes to thank the CFP Th\'eorie-IN2P3 for financial support.
  \end{acknowledgments}

\appendix 

\bibliography{mybib}

\providecommand{\href}[2]{#2}\begingroup\raggedright\begin{thebibliography}{100}

\bibitem{1987ARA&A..25..425T}
V.~{Trimble}, {\it {Existence and nature of dark matter in the universe}},
  {\em \araa} {\bf 25} (1987) 425--472.

\bibitem{1991ApJ...376...51W}
T.~P. {Walker}, G.~{Steigman}, H.-S. {Kang}, D.~M. {Schramm}, and K.~A.
  {Olive}, {\it {Primordial nucleosynthesis redux}},  {\em \apj} {\bf 376}
  (July, 1991) 51--69.

\bibitem{2011ApJS..192...18K}
E.~{Komatsu}, K.~M. {Smith}, J.~{Dunkley}, {\em et.~al.}, {\it {Seven-year
  Wilkinson Microwave Anisotropy Probe (WMAP) Observations: Cosmological
  Interpretation}},  {\em \apjs} {\bf 192} (Feb., 2011) 18,
  [\href{http://xxx.lanl.gov/abs/1001.4538}{{\tt arXiv:1001.4538}}].

\bibitem{1988ARNPS..38..751P}
J.~R. {Primack}, D.~{Seckel}, and B.~{Sadoulet}, {\it {Detection of cosmic dark
  matter}},  {\em Annual Review of Nuclear and Particle Science} {\bf 38}
  (1988) 751--807.

\bibitem{1996PhR...267..195J}
G.~{Jungman}, M.~{Kamionkowski}, and K.~{Griest}, {\it {Supersymmetric dark
  matter}},  {\em \physrep} {\bf 267} (Mar., 1996) 195--373,
  [\href{http://xxx.lanl.gov/abs/hep-ph/9506380}{{\tt hep-ph/9506380}}].

\bibitem{2009NJPh...11j5029P}
J.~R. {Primack}, {\it {Cosmology: small-scale issues}},  {\em New Journal of
  Physics} {\bf 11} (Oct., 2009) 105029,
  [\href{http://xxx.lanl.gov/abs/0909.2247}{{\tt arXiv:0909.2247}}].

\bibitem{1993ApJ...411..439S}
J.~{Silk} and A.~{Stebbins}, {\it {Clumpy cold dark matter}},  {\em \apj} {\bf
  411} (July, 1993) 439--449.

\bibitem{1978ApJ...223.1015G}
J.~E. {Gunn}, B.~W. {Lee}, I.~{Lerche}, D.~N. {Schramm}, and G.~{Steigman},
  {\it {Some astrophysical consequences of the existence of a heavy stable
  neutral lepton}},  {\em \apj} {\bf 223} (Aug., 1978) 1015--1031.

\bibitem{1984PhRvL..53..624S}
J.~{Silk} and M.~{Srednicki}, {\it {Cosmic-ray antiprotons as a probe of a
  photino-dominated universe}},  {\em Physical Review Letters} {\bf 53} (Aug.,
  1984) 624--627.

\bibitem{1985ApJ...299.1001K}
L.~M. {Krauss}, K.~{Freese}, D.~N. {Spergel}, and W.~H. {Press}, {\it {Cold
  dark matter candidates and the solar neutrino problem}},  {\em \apj} {\bf
  299} (Dec., 1985) 1001--1006.

\bibitem{2009ApJ...697.1071A}
W.~B. {Atwood} {\em et.~al.}, {\it {The Large Area Telescope on the Fermi
  Gamma-Ray Space Telescope Mission}},  {\em \apj} {\bf 697} (June, 2009)
  1071--1102, [\href{http://xxx.lanl.gov/abs/0902.1089}{{\tt
  arXiv:0902.1089}}].

\bibitem{2012JCAP...07..054B}
T.~{Bringmann}, X.~{Huang}, A.~{Ibarra}, S.~{Vogl}, and C.~{Weniger}, {\it
  {Fermi LAT search for internal bremsstrahlung signatures from dark matter
  annihilation}},  {\em \jcap} {\bf 7} (July, 2012) 54,
  [\href{http://xxx.lanl.gov/abs/1203.1312}{{\tt arXiv:1203.1312}}].

\bibitem{2012arXiv1204.2797W}
C.~{Weniger}, {\it {A Tentative Gamma-Ray Line from Dark Matter Annihilation at
  the Fermi Large Area Telescope}},  {\em ArXiv e-prints} (Apr., 2012)
  [\href{http://xxx.lanl.gov/abs/1204.2797}{{\tt arXiv:1204.2797}}].

\bibitem{2009NuPhB.821..399C}
M.~{Cirelli} and P.~{Panci}, {\it {Inverse Compton constraints on the Dark
  Matter $e^{\pm}$ excesses}},  {\em Nuclear Physics B} {\bf 821} (Nov., 2009)
  399--416, [\href{http://xxx.lanl.gov/abs/0904.3830}{{\tt arXiv:0904.3830}}].

\bibitem{2010NuPhB.831..178M}
P.~{Meade}, M.~{Papucci}, A.~{Strumia}, and T.~{Volansky}, {\it {Dark Matter
  interpretations of the e$^{±}$ excesses after FERMI}},  {\em Nuclear Physics
  B} {\bf 831} (May, 2010) 178--203,
  [\href{http://xxx.lanl.gov/abs/0905.0480}{{\tt arXiv:0905.0480}}].

\bibitem{2010JCAP...03..014P}
M.~{Papucci} and A.~{Strumia}, {\it {Robust implications on dark matter from
  the first FERMI sky {$\gamma$} map}},  {\em \jcap} {\bf 3} (Mar., 2010) 14,
  [\href{http://xxx.lanl.gov/abs/0912.0742}{{\tt arXiv:0912.0742}}].

\bibitem{2010NuPhB.840..284C}
M.~{Cirelli}, P.~{Panci}, and P.~D. {Serpico}, {\it {Diffuse gamma ray
  constraints on annihilating or decaying Dark Matter after Fermi}},  {\em
  Nuclear Physics B} {\bf 840} (Nov., 2010) 284--303,
  [\href{http://xxx.lanl.gov/abs/0912.0663}{{\tt arXiv:0912.0663}}].

\bibitem{2010JCAP...04..014A}
A.~A. {Abdo} {\em et.~al.}, {\it {Constraints on cosmological dark matter
  annihilation from the Fermi-LAT isotropic diffuse gamma-ray measurement}},
  {\em \jcap} {\bf 4} (Apr., 2010) 14,
  [\href{http://xxx.lanl.gov/abs/1002.4415}{{\tt arXiv:1002.4415}}].

\bibitem{2010JCAP...11..041A}
K.~N. {Abazajian}, P.~{Agrawal}, Z.~{Chacko}, and C.~{Kilic}, {\it
  {Conservative constraints on dark matter from the Fermi-LAT isotropic diffuse
  gamma-ray background spectrum}},  {\em \jcap} {\bf 11} (Nov., 2010) 41,
  [\href{http://xxx.lanl.gov/abs/1002.3820}{{\tt arXiv:1002.3820}}].

\bibitem{2010JCAP...07..008H}
G.~{H{\"u}tsi}, A.~{Hektor}, and M.~{Raidal}, {\it {Implications of the
  Fermi-LAT diffuse gamma-ray measurements on annihilating or decaying dark
  matter}},  {\em \jcap} {\bf 7} (July, 2010) 8,
  [\href{http://xxx.lanl.gov/abs/1004.2036}{{\tt arXiv:1004.2036}}].

\bibitem{2012PhRvD..85d3509A}
K.~N. {Abazajian}, S.~{Blanchet}, and J.~P. {Harding}, {\it {Current and future
  constraints on dark matter from prompt and inverse-Compton photon emission in
  the isotropic diffuse gamma-ray background}},  {\em \prd} {\bf 85} (Feb.,
  2012) 043509, [\href{http://xxx.lanl.gov/abs/1011.5090}{{\tt
  arXiv:1011.5090}}].

\bibitem{2011JCAP...03..051C}
M.~{Cirelli}, G.~{Corcella}, A.~{Hektor}, G.~{H{\"u}tsi}, M.~{Kadastik},
  P.~{Panci}, M.~{Raidal}, F.~{Sala}, and A.~{Strumia}, {\it {PPPC 4 DM ID: a
  poor particle physicist cookbook for dark matter indirect detection}},  {\em
  \jcap} {\bf 3} (Mar., 2011) 51,
  [\href{http://xxx.lanl.gov/abs/1012.4515}{{\tt arXiv:1012.4515}}].

\bibitem{2012JCAP...01..041A}
K.~N. {Abazajian} and J.~P. {Harding}, {\it {Constraints on WIMP and
  Sommerfeld-enhanced dark matter annihilation from HESS observations of the
  galactic center}},  {\em \jcap} {\bf 1} (Jan., 2012) 41,
  [\href{http://xxx.lanl.gov/abs/1110.6151}{{\tt arXiv:1110.6151}}].

\bibitem{2011arXiv1111.2835A}
K.~N. {Abazajian}, P.~{Agrawal}, Z.~{Chacko}, and C.~{Kilic}, {\it {Lower
  Limits on the Strengths of Gamma Ray Lines from WIMP Dark Matter
  Annihilation}},  {\em ArXiv e-prints} (Nov., 2011)
  [\href{http://xxx.lanl.gov/abs/1111.2835}{{\tt arXiv:1111.2835}}].

\bibitem{2010PhRvL.104j1101A}
A.~A. {Abdo} {\em et.~al.}, {\it {Spectrum of the Isotropic Diffuse Gamma-Ray
  Emission Derived from First-Year Fermi Large Area Telescope Data}},  {\em
  Physical Review Letters} {\bf 104} (Mar., 2010) 101101,
  [\href{http://xxx.lanl.gov/abs/1002.3603}{{\tt arXiv:1002.3603}}].

\bibitem{2011PhRvD..83b3518P}
L.~{Pieri}, J.~{Lavalle}, G.~{Bertone}, and E.~{Branchini}, {\it {Implications
  of high-resolution simulations on indirect dark matter searches}},  {\em
  \prd} {\bf 83} (Jan., 2011) 023518--+,
  [\href{http://xxx.lanl.gov/abs/0908.0195}{{\tt arXiv:0908.0195}}].

\bibitem{1974ApJ...187..425P}
W.~H. {Press} and P.~{Schechter}, {\it {Formation of Galaxies and Clusters of
  Galaxies by Self-Similar Gravitational Condensation}},  {\em \apj} {\bf 187}
  (Feb., 1974) 425--438.

\bibitem{2001MNRAS.323....1S}
R.~K. {Sheth}, H.~J. {Mo}, and G.~{Tormen}, {\it {Ellipsoidal collapse and an
  improved model for the number and spatial distribution of dark matter
  haloes}},  {\em \mnras} {\bf 323} (May, 2001) 1--12,
  [\href{http://xxx.lanl.gov/abs/astro-ph/9907024}{{\tt astro-ph/9907024}}].

\bibitem{2008Natur.454..735D}
J.~{Diemand}, M.~{Kuhlen}, P.~{Madau}, M.~{Zemp}, B.~{Moore}, D.~{Potter}, and
  J.~{Stadel}, {\it {Clumps and streams in the local dark matter
  distribution}},  {\em \nat} {\bf 454} (Aug., 2008) 735--738,
  [\href{http://xxx.lanl.gov/abs/0805.1244}{{\tt arXiv:0805.1244}}].

\bibitem{2008ApJ...679.1260M}
P.~{Madau}, J.~{Diemand}, and M.~{Kuhlen}, {\it {Dark Matter Subhalos and the
  Dwarf Satellites of the Milky Way}},  {\em \apj} {\bf 679} (June, 2008)
  1260--1271, [\href{http://xxx.lanl.gov/abs/0802.2265}{{\tt
  arXiv:0802.2265}}].

\bibitem{2008MNRAS.391.1685S}
V.~{Springel}, J.~{Wang}, M.~{Vogelsberger}, A.~{Ludlow}, A.~{Jenkins},
  A.~{Helmi}, J.~F. {Navarro}, C.~S. {Frenk}, and S.~D.~M. {White}, {\it {The
  Aquarius Project: the subhaloes of galactic haloes}},  {\em \mnras} {\bf 391}
  (Dec., 2008) 1685--1711, [\href{http://xxx.lanl.gov/abs/0809.0898}{{\tt
  arXiv:0809.0898}}].

\bibitem{2009NJPh...11j5027B}
T.~{Bringmann}, {\it {Particle models and the small-scale structure of dark
  matter}},  {\em New Journal of Physics} {\bf 11} (Oct., 2009) 105027--+,
  [\href{http://xxx.lanl.gov/abs/0903.0189}{{\tt arXiv:0903.0189}}].

\bibitem{2005Natur.433..389D}
J.~{Diemand}, B.~{Moore}, and J.~{Stadel}, {\it {Earth-mass dark-matter haloes
  as the first structures in the early Universe}},  {\em \nat} {\bf 433} (Jan.,
  2005) 389--391, [\href{http://xxx.lanl.gov/abs/astro-ph/0501589}{{\tt
  astro-ph/0501589}}].

\bibitem{1999PhRvD..59d3506B}
L.~{Bergstr{\"o}m}, J.~{Edsj{\"o}}, P.~{Gondolo}, and P.~{Ullio}, {\it {Clumpy
  neutralino dark matter}},  {\em \prd} {\bf 59} (Feb., 1999) 043506,
  [\href{http://xxx.lanl.gov/abs/astro-ph/9806072}{{\tt astro-ph/9806072}}].

\bibitem{2003PhRvD..68j3003B}
V.~{Berezinsky}, V.~{Dokuchaev}, and Y.~{Eroshenko}, {\it {Small-scale clumps
  in the galactic halo and dark matter annihilation}},  {\em \prd} {\bf 68}
  (Nov., 2003) 103003--+, [\href{http://xxx.lanl.gov/abs/astro-ph/0301551}{{\tt
  astro-ph/0301551}}].

\bibitem{2007A&A...462..827L}
J.~{Lavalle}, J.~{Pochon}, P.~{Salati}, and R.~{Taillet}, {\it {Clumpiness of
  dark matter and the positron annihilation signal}},  {\em \aap} {\bf 462}
  (Feb., 2007) 827--840, [\href{http://xxx.lanl.gov/abs/astro-ph/0603796}{{\tt
  astro-ph/0603796}}].

\bibitem{2002PhRvD..66l3502U}
P.~{Ullio}, L.~{Bergstr{\"o}m}, J.~{Edsj{\"o}}, and C.~{Lacey}, {\it
  {Cosmological dark matter annihilations into {$\gamma$} rays: A closer
  look}},  {\em \prd} {\bf 66} (Dec., 2002) 123502--+,
  [\href{http://xxx.lanl.gov/abs/astro-ph/0207125}{{\tt astro-ph/0207125}}].

\bibitem{2012ApJ...750....3A}
M.~{Ackermann} {\em et.~al.}, {\it {Fermi-LAT Observations of the Diffuse
  {$\gamma$}-Ray Emission: Implications for Cosmic Rays and the Interstellar
  Medium}},  {\em \apj} {\bf 750} (May, 2012) 3,
  [\href{http://xxx.lanl.gov/abs/1202.4039}{{\tt arXiv:1202.4039}}].

\bibitem{1997ApJ...490..493N}
J.~F. {Navarro}, C.~S. {Frenk}, and S.~D.~M. {White}, {\it {A Universal Density
  Profile from Hierarchical Clustering}},  {\em \apj} {\bf 490} (Dec., 1997)
  493--+, [\href{http://xxx.lanl.gov/abs/astro-ph/9611107}{{\tt
  astro-ph/9611107}}].

\bibitem{2011MNRAS.414.2446M}
P.~J. {McMillan}, {\it {Mass models of the Milky Way}},  {\em \mnras} {\bf 414}
  (July, 2011) 2446--2457, [\href{http://xxx.lanl.gov/abs/1102.4340}{{\tt
  arXiv:1102.4340}}].

\bibitem{2010JCAP...08..004C}
R.~{Catena} and P.~{Ullio}, {\it {A novel determination of the local dark
  matter density}},  {\em \jcap} {\bf 8} (Aug., 2010) 4--+,
  [\href{http://xxx.lanl.gov/abs/0907.0018}{{\tt arXiv:0907.0018}}].

\bibitem{2010A&A...523A..83S}
P.~{Salucci}, F.~{Nesti}, G.~{Gentile}, and C.~{Frigerio Martins}, {\it {The
  dark matter density at the Sun's location}},  {\em \aap} {\bf 523} (Nov.,
  2010) A83+, [\href{http://xxx.lanl.gov/abs/1003.3101}{{\tt
  arXiv:1003.3101}}].

\bibitem{2012arXiv1205.4033B}
J.~{Bovy} and S.~{Tremaine}, {\it {On the local dark matter density}},  {\em
  ArXiv e-prints} (May, 2012) [\href{http://xxx.lanl.gov/abs/1205.4033}{{\tt
  arXiv:1205.4033}}].

\bibitem{2005ICRC....4...77P}
T.~A. {Porter} and {et al.}, {\it {A new estimate of the Galactic interstellar
  radiation field between 0.1um and 1000um}},  in {\em International Cosmic Ray
  Conference}, vol.~4 of {\em International Cosmic Ray Conference}, pp.~77--+,
  2005.

\bibitem{2011ApJ...727...38S}
T.~{Shibata}, T.~{Ishikawa}, and S.~{Sekiguchi}, {\it {A Possible Approach to
  Three-dimensional Cosmic-ray Propagation in the Galaxy. IV. Electrons and
  Electron-induced {$\gamma$}-rays}},  {\em \apj} {\bf 727} (Jan., 2011) 38--+,
  [\href{http://xxx.lanl.gov/abs/1010.5652}{{\tt arXiv:1010.5652}}].

\bibitem{2000ApJ...537..763S}
A.~W. {Strong}, I.~V. {Moskalenko}, and O.~{Reimer}, {\it {Diffuse Continuum
  Gamma Rays from the Galaxy}},  {\em \apj} {\bf 537} (July, 2000) 763--784,
  [\href{http://xxx.lanl.gov/abs/astro-ph/9811296}{{\tt astro-ph/9811296}}].

\bibitem{2005PhRvL..94q1301B}
J.~F. {Beacom}, N.~F. {Bell}, and G.~{Bertone}, {\it {Gamma-Ray Constraint on
  Galactic Positron Production by MeV Dark Matter}},  {\em Physical Review
  Letters} {\bf 94} (May, 2005) 171301--+,
  [\href{http://xxx.lanl.gov/abs/astro-ph/0409403}{{\tt astro-ph/0409403}}].

\bibitem{2009JCAP...02..021I}
A.~{Ibarra} and D.~{Tran}, {\it {Decaying dark matter and the PAMELA anomaly}},
   {\em \jcap} {\bf 2} (Feb., 2009) 21,
  [\href{http://xxx.lanl.gov/abs/0811.1555}{{\tt arXiv:0811.1555}}].

\bibitem{2010JCAP...01..009I}
A.~{Ibarra}, D.~{Tran}, and C.~{Weniger}, {\it {Decaying dark matter in light
  of the PAMELA and Fermi LAT data}},  {\em \jcap} {\bf 1} (Jan., 2010) 9,
  [\href{http://xxx.lanl.gov/abs/0906.1571}{{\tt arXiv:0906.1571}}].

\bibitem{2012PhRvD..85b3004C}
F.~{Calore}, V.~{de Romeri}, and F.~{Donato}, {\it {Conservative upper limits
  on WIMP annihilation cross section from Fermi-LAT {$\gamma$} rays}},  {\em
  \prd} {\bf 85} (Jan., 2012) 023004,
  [\href{http://xxx.lanl.gov/abs/1105.4230}{{\tt arXiv:1105.4230}}].

\bibitem{2009Natur.458..607A}
O.~{Adriani} {\em et.~al.}, {\it {An anomalous positron abundance in cosmic
  rays with energies 1.5-100GeV}},  {\em \nat} {\bf 458} (Apr., 2009) 607--609,
  [\href{http://xxx.lanl.gov/abs/0810.4995}{{\tt arXiv:0810.4995}}].

\bibitem{2012PhRvL.108a1103A}
M.~{Ackermann} {\em et.~al.}, {\it {Measurement of Separate Cosmic-Ray Electron
  and Positron Spectra with the Fermi Large Area Telescope}},  {\em Physical
  Review Letters} {\bf 108} (Jan., 2012) 011103,
  [\href{http://xxx.lanl.gov/abs/1109.0521}{{\tt arXiv:1109.0521}}].

\bibitem{2009PhRvL.102r1101A}
A.~A. {Abdo} {\em et.~al.}, {\it {Measurement of the Cosmic Ray $e^{+}+e^{-}$
  Spectrum from 20GeV to 1TeV with the Fermi Large Area Telescope}},  {\em
  Physical Review Letters} {\bf 102} (May, 2009) 181101--+,
  [\href{http://xxx.lanl.gov/abs/0905.0025}{{\tt arXiv:0905.0025}}].

\bibitem{1965PhRv..137.1306J}
F.~C. {Jones}, {\it {Inverse Compton Scattering of Cosmic-Ray Electrons}},
  {\em Physical Review} {\bf 137} (Mar., 1965) 1306--1311.

\bibitem{1968PhRv..167.1159J}
F.~C. {Jones}, {\it {Calculated Spectrum of Inverse-Compton-Scattered
  Photons}},  {\em Physical Review} {\bf 167} (Mar., 1968) 1159--1169.

\bibitem{1970RvMP...42..237B}
G.~R. {Blumenthal} and R.~J. {Gould}, {\it {Bremsstrahlung, Synchrotron
  Radiation, and Compton Scattering of High-Energy Electrons Traversing Dilute
  Gases}},  {\em Reviews of Modern Physics} {\bf 42} (1970) 237--271.

\bibitem{2006JHEP...05..026S}
T.~{Sj{\"o}strand}, S.~{Mrenna}, and P.~{Skands}, {\it {PYTHIA 6.4 physics and
  manual}},  {\em Journal of High Energy Physics} {\bf 5} (May, 2006) 26--+,
  [\href{http://xxx.lanl.gov/abs/hep-ph/0603175}{{\tt hep-ph/0603175}}].

\bibitem{1959SvA.....3...22S}
S.~I. {Syrovatskii}, {\it {The Distribution of Relativistic Electrons in the
  Galaxy and the Spectrum of Synchrotron Radio Emission.}},  {\em \sovast} {\bf
  3} (Feb., 1959) 22.

\bibitem{1964ocr..book.....G}
V.~L. {Ginzburg} and S.~I. {Syrovatskii}, {\em {The Origin of Cosmic Rays}}.
\newblock New York: Macmillan, 1964.

\bibitem{berezinsky_book_90}
V.~S. {Berezinskii}, S.~V. {Bulanov}, V.~A. {Dogiel}, and V.~S. {Ptuskin}, {\em
  {Astrophysics of cosmic rays}}.
\newblock Amsterdam: North-Holland, edited by Ginzburg, V.L., 1990.

\bibitem{2009A&A...501..821D}
T.~{Delahaye}, F.~{Donato}, N.~{Fornengo}, J.~{Lavalle}, R.~{Lineros},
  P.~{Salati}, and R.~{Taillet}, {\it {Galactic secondary positron flux at the
  Earth}},  {\em \aap} {\bf 501} (July, 2009) 821--833,
  [\href{http://xxx.lanl.gov/abs/0809.5268}{{\tt arXiv:0809.5268}}].

\bibitem{1998ApJ...509..212S}
A.~W. {Strong} and I.~V. {Moskalenko}, {\it {Propagation of Cosmic-Ray Nucleons
  in the Galaxy}},  {\em \apj} {\bf 509} (Dec., 1998) 212--228,
  [\href{http://xxx.lanl.gov/abs/astro-ph/9807150}{{\tt astro-ph/9807150}}].

\bibitem{2009arXiv0909.4548D}
G.~{Di Bernardo}, C.~{Evoli}, D.~{Gaggero}, D.~{Grasso}, and L.~{Maccione},
  {\it {Unified interpretation of cosmic-ray nuclei and antiproton recent
  measurements}},  {\em ArXiv e-prints} (Sept., 2009)
  [\href{http://xxx.lanl.gov/abs/0909.4548}{{\tt arXiv:0909.4548}}].

\bibitem{1974Ap&SS..29..305B}
S.~V. {Bulanov} and V.~A. {Dogel}, {\it {The Influence of the Energy Dependence
  of the Diffusion Coefficient on the Spectrum of the Electron Component of
  Cosmic Rays and the Radio Background Radiation of the Galaxy}},  {\em \apss}
  {\bf 29} (Aug., 1974) 305--318.

\bibitem{1998PhRvD..59b3511B}
E.~A. {Baltz} and J.~{Edsj{\"o}}, {\it {Positron propagation and fluxes from
  neutralino annihilation in the halo}},  {\em \prd} {\bf 59} (Jan., 1998)
  023511, [\href{http://xxx.lanl.gov/abs/astro-ph/9808243}{{\tt
  astro-ph/9808243}}].

\bibitem{2008PhRvD..77f3527D}
T.~{Delahaye}, R.~{Lineros}, F.~{Donato}, N.~{Fornengo}, and P.~{Salati}, {\it
  {Positrons from dark matter annihilation in the galactic halo: Theoretical
  uncertainties}},  {\em \prd} {\bf 77} (Mar., 2008) 063527--+,
  [\href{http://xxx.lanl.gov/abs/0712.2312}{{\tt arXiv:0712.2312}}].

\bibitem{2012arXiv1205.1004L}
J.~{Lavalle} and P.~{Salati}, {\it {Dark Matter Indirect Signatures}},  {\em
  ArXiv e-prints} (May, 2012) [\href{http://xxx.lanl.gov/abs/1205.1004}{{\tt
  arXiv:1205.1004}}].

\bibitem{2010A&A...524A..51D}
T.~{Delahaye}, J.~{Lavalle}, R.~{Lineros}, F.~{Donato}, and N.~{Fornengo}, {\it
  {Galactic electrons and positrons at the Earth: new estimate of the primary
  and secondary fluxes}},  {\em \aap} {\bf 524} (Dec., 2010) A51+,
  [\href{http://xxx.lanl.gov/abs/1002.1910}{{\tt arXiv:1002.1910}}].

\bibitem{2002PhRvD..65b3002C}
F.~{Casse}, M.~{Lemoine}, and G.~{Pelletier}, {\it {Transport of cosmic rays in
  chaotic magnetic fields}},  {\em \prd} {\bf 65} (Jan., 2002) 023002,
  [\href{http://xxx.lanl.gov/abs/astro-ph/0109223}{{\tt astro-ph/0109223}}].

\bibitem{2009ncrd.book.....S}
A.~{Shalchi}, {\em {Nonlinear Cosmic Ray Diffusion Theories}}.
\newblock {Springer}, 2009.

\bibitem{2004PhRvD..69f3501D}
F.~{Donato}, N.~{Fornengo}, D.~{Maurin}, P.~{Salati}, and R.~{Taillet}, {\it
  {Antiprotons in cosmic rays from neutralino annihilation}},  {\em \prd} {\bf
  69} (Mar., 2004) 063501--+,
  [\href{http://xxx.lanl.gov/abs/astro-ph/0306207}{{\tt astro-ph/0306207}}].

\bibitem{2008A&A...479..427L}
J.~{Lavalle}, Q.~{Yuan}, D.~{Maurin}, and X.-J. {Bi}, {\it {Full calculation of
  clumpiness boost factors for antimatter cosmic rays in the light of
  {$\Lambda$}CDM N-body simulation results. Abandoning hope in clumpiness
  enhancement?}},  {\em \aap} {\bf 479} (Feb., 2008) 427--452,
  [\href{http://xxx.lanl.gov/abs/0709.3634}{{\tt arXiv:0709.3634}}].

\bibitem{2010PhRvD..82h1302L}
J.~{Lavalle}, {\it {10 GeV dark matter candidates and cosmic-ray antiprotons}},
   {\em \prd} {\bf 82} (Oct., 2010) 081302--+,
  [\href{http://xxx.lanl.gov/abs/1007.5253}{{\tt arXiv:1007.5253}}].

\bibitem{2001ApJ...555..585M}
D.~{Maurin}, F.~{Donato}, R.~{Taillet}, and P.~{Salati}, {\it {Cosmic Rays
  below Z=30 in a Diffusion Model: New Constraints on Propagation Parameters}},
   {\em \apj} {\bf 555} (July, 2001) 585--596,
  [\href{http://xxx.lanl.gov/abs/astro-ph/0101231}{{\tt astro-ph/0101231}}].

\bibitem{2010A&A...516A..66P}
A.~{Putze}, L.~{Derome}, and D.~{Maurin}, {\it {A Markov Chain Monte Carlo
  technique to sample transport and source parameters of Galactic cosmic rays.
  II. Results for the diffusion model combining B/C and radioactive nuclei}},
  {\em \aap} {\bf 516} (June, 2010) A66+,
  [\href{http://xxx.lanl.gov/abs/1001.0551}{{\tt arXiv:1001.0551}}].

\bibitem{2009PhRvL.103y1101A}
A.~A. {Abdo} {\em et.~al.}, {\it {Fermi Large Area Telescope Measurements of
  the Diffuse Gamma-Ray Emission at Intermediate Galactic Latitudes}},  {\em
  Physical Review Letters} {\bf 103} (Dec., 2009) 251101--+.

\bibitem{2011MNRAS.414..985L}
J.~{Lavalle}, {\it {Impact of the spectral hardening of TeV cosmic rays on the
  prediction of the secondary positron flux}},  {\em \mnras} {\bf 414} (June,
  2011) 985--991, [\href{http://xxx.lanl.gov/abs/1011.3063}{{\tt
  arXiv:1011.3063}}].

\bibitem{2010PhRvD..81j3521K}
M.~D. {Kistler} and J.~M. {Siegal-Gaskins}, {\it {Gamma-ray signatures of
  annihilation to charged leptons in dark matter substructure}},  {\em \prd}
  {\bf 81} (May, 2010) 103521, [\href{http://xxx.lanl.gov/abs/0909.0519}{{\tt
  arXiv:0909.0519}}].

\bibitem{2010PhRvD..82h3521L}
J.~{Lavalle}, {\it {Sunyaev-Zel'dovich effects from annihilating dark matter in
  the Milky Way: Smooth halo, subhalos, and intermediate-mass black holes}},
  {\em \prd} {\bf 82} (Oct., 2010) 083521--+,
  [\href{http://xxx.lanl.gov/abs/1008.5124}{{\tt arXiv:1008.5124}}].

\bibitem{2002A&A...388..676B}
A.~{Barrau}, G.~{Boudoul}, F.~{Donato}, D.~{Maurin}, P.~{Salati}, and
  R.~{Taillet}, {\it {Antiprotons from primordial black holes}},  {\em \aap}
  {\bf 388} (June, 2002) 676--687,
  [\href{http://xxx.lanl.gov/abs/astro-ph/0112486}{{\tt astro-ph/0112486}}].

\bibitem{2010PhRvD..82d3505P}
M.~{Perelstein} and B.~{Shakya}, {\it {Remarks on calculation of positron flux
  from galactic dark matter}},  {\em \prd} {\bf 82} (Aug., 2010) 043505,
  [\href{http://xxx.lanl.gov/abs/1002.4588}{{\tt arXiv:1002.4588}}].

\bibitem{2011PhRvD..83l3508P}
M.~{Perelstein} and B.~{Shakya}, {\it {Antiprotons from dark matter: Effects of
  a position-dependent diffusion coefficient}},  {\em \prd} {\bf 83} (June,
  2011) 123508, [\href{http://xxx.lanl.gov/abs/1012.3772}{{\tt
  arXiv:1012.3772}}].

\bibitem{1981Ap&SS..79..321A}
F.~A. {Aharonian} and A.~M. {Atoyan}, {\it {Compton scattering of relativistic
  electrons in compact X-ray sources}},  {\em \apss} {\bf 79} (Oct., 1981)
  321--336.

\bibitem{2000ApJ...528..357M}
I.~V. {Moskalenko} and A.~W. {Strong}, {\it {Anisotropic Inverse Compton
  Scattering in the Galaxy}},  {\em \apj} {\bf 528} (Jan., 2000) 357--367,
  [\href{http://xxx.lanl.gov/abs/astro-ph/9811284}{{\tt astro-ph/9811284}}].

\bibitem{2011A&A...534A..54S}
A.~W. {Strong}, E.~{Orlando}, and T.~R. {Jaffe}, {\it {The interstellar
  cosmic-ray electron spectrum from synchrotron radiation and direct
  measurements}},  {\em \aap} {\bf 534} (Oct., 2011) A54,
  [\href{http://xxx.lanl.gov/abs/1108.4822}{{\tt arXiv:1108.4822}}].

\bibitem{2012JCAP...01..049B}
T.~{Bringmann}, F.~{Donato}, and R.~A. {Lineros}, {\it {Radio data and
  synchrotron emission in consistent cosmic ray models}},  {\em \jcap} {\bf 1}
  (Jan., 2012) 49, [\href{http://xxx.lanl.gov/abs/1106.4821}{{\tt
  arXiv:1106.4821}}].

\bibitem{2006NuPhB.741...83B}
X.-J. {Bi}, {\it {Gamma rays from the neutralino dark matter annihilations in
  the Milky Way substructures}},  {\em Nuclear Physics B} {\bf 741} (May, 2006)
  83--107, [\href{http://xxx.lanl.gov/abs/astro-ph/0510714}{{\tt
  astro-ph/0510714}}].

\bibitem{2010arXiv1008.1801Z}
L.~{Zhang}, F.~{Miniati}, and G.~{Sigl}, {\it {Inverse Compton gamma-rays from
  Galactic dark matter annihilation: Anisotropy signatures}},  {\em ArXiv
  e-prints} (Aug., 2010) [\href{http://xxx.lanl.gov/abs/1008.1801}{{\tt
  arXiv:1008.1801}}].

\bibitem{2007ApJ...667..859D}
J.~{Diemand}, M.~{Kuhlen}, and P.~{Madau}, {\it {Formation and Evolution of
  Galaxy Dark Matter Halos and Their Substructure}},  {\em \apj} {\bf 667}
  (Oct., 2007) 859--877, [\href{http://xxx.lanl.gov/abs/astro-ph/0703337}{{\tt
  astro-ph/0703337}}].

\bibitem{2008MNRAS.384.1627P}
L.~{Pieri}, G.~{Bertone}, and E.~{Branchini}, {\it {Dark matter annihilation in
  substructures revised}},  {\em \mnras} {\bf 384} (Mar., 2008) 1627--1637,
  [\href{http://xxx.lanl.gov/abs/0706.2101}{{\tt arXiv:0706.2101}}].

\bibitem{2010PhRvD..81d3532K}
M.~{Kamionkowski}, S.~M. {Koushiappas}, and M.~{Kuhlen}, {\it {Galactic
  substructure and dark-matter annihilation in the Milky Way halo}},  {\em
  \prd} {\bf 81} (Feb., 2010) 043532,
  [\href{http://xxx.lanl.gov/abs/1001.3144}{{\tt arXiv:1001.3144}}].

\bibitem{2012CoPhC.183..656C}
A.~{Charbonnier}, C.~{Combet}, and D.~{Maurin}, {\it {CLUMPY: A code for
  {$\gamma$}-ray signals from dark matter structures}},  {\em Computer Physics
  Communications} {\bf 183} (Mar., 2012) 656--668,
  [\href{http://xxx.lanl.gov/abs/1201.4728}{{\tt arXiv:1201.4728}}].

\bibitem{Morselli}
{\bf Fermi-LAT} Collaboration, A.~Morselli, {\it {Fermi results}},  {\em Dark
  Side of the Universe conference, Beijing} (2011).

\bibitem{2012NuPhB.854..738C}
D.~G. {Cerde{\~n}o}, T.~{Delahaye}, and J.~{Lavalle}, {\it {Cosmic-ray
  antiproton constraints on light singlino-like dark matter candidates}},  {\em
  Nuclear Physics B} {\bf 854} (Jan., 2012) 738--779,
  [\href{http://xxx.lanl.gov/abs/1108.1128}{{\tt arXiv:1108.1128}}].

\bibitem{2012PhRvD..86b3506S}
G.~{Steigman}, B.~{Dasgupta}, and J.~F. {Beacom}, {\it {Precise relic WIMP
  abundance and its impact on searches for dark matter annihilation}},  {\em
  \prd} {\bf 86} (July, 2012) 023506,
  [\href{http://xxx.lanl.gov/abs/1204.3622}{{\tt arXiv:1204.3622}}].

\bibitem{1999ApJ...526..215B}
L.~{Bergstr{\"o}m}, J.~{Edsj{\"o}}, and P.~{Ullio}, {\it {Cosmic Antiprotons as
  a Probe for Supersymmetric Dark Matter?}},  {\em \apj} {\bf 526} (Nov., 1999)
  215--235, [\href{http://xxx.lanl.gov/abs/astro-ph/9902012}{{\tt
  astro-ph/9902012}}].

\bibitem{2011PhRvL.107x1302A}
M.~{Ackermann} {\em et.~al.}, {\it {Constraining Dark Matter Models from a
  Combined Analysis of Milky Way Satellites with the Fermi Large Area
  Telescope}},  {\em Physical Review Letters} {\bf 107} (Dec., 2011) 241302,
  [\href{http://xxx.lanl.gov/abs/1108.3546}{{\tt arXiv:1108.3546}}].

\bibitem{2011PhRvL.107x1303G}
A.~{Geringer-Sameth} and S.~M. {Koushiappas}, {\it {Exclusion of Canonical
  Weakly Interacting Massive Particles by Joint Analysis of Milky Way Dwarf
  Galaxies with Data from the Fermi Gamma-Ray Space Telescope}},  {\em Physical
  Review Letters} {\bf 107} (Dec., 2011) 241303,
  [\href{http://xxx.lanl.gov/abs/1108.2914}{{\tt arXiv:1108.2914}}].

\bibitem{2009ApJ...702..523I}
Y.~{Inoue} and T.~{Totani}, {\it {The Blazar Sequence and the Cosmic Gamma-ray
  Background Radiation in the Fermi Era}},  {\em \apj} {\bf 702} (Sept., 2009)
  523--536, [\href{http://xxx.lanl.gov/abs/0810.3580}{{\tt arXiv:0810.3580}}].

\bibitem{2011ApJ...728...73I}
Y.~{Inoue} and T.~{Totani}, {\it {ERRATUM: ''The Blazar Sequence and the Cosmic
  Gamma-ray Background Radiation in the Fermi Era'' <A
  href=''/abs/2009ApJ...702..523I''>(2009, ApJ, 702, 523)</A>}},  {\em \apj}
  {\bf 728} (Feb., 2011) 73.

\bibitem{2011PhRvD..84j3007A}
K.~N. {Abazajian}, S.~{Blanchet}, and J.~P. {Harding}, {\it {Contribution of
  blazars to the extragalactic diffuse gamma-ray background and their future
  spatial resolution}},  {\em \prd} {\bf 84} (Nov., 2011) 103007,
  [\href{http://xxx.lanl.gov/abs/1012.1247}{{\tt arXiv:1012.1247}}].

\bibitem{2008ApJ...672L...5I}
Y.~{Inoue}, T.~{Totani}, and Y.~{Ueda}, {\it {The Cosmic MeV Gamma-Ray
  Background and Hard X-Ray Spectra of Active Galactic Nuclei: Implications for
  the Origin of Hot AGN Coronae}},  {\em \apjl} {\bf 672} (Jan., 2008) L5--L8,
  [\href{http://xxx.lanl.gov/abs/0709.3877}{{\tt arXiv:0709.3877}}].

\bibitem{2010ApJ...722L.199F}
B.~D. {Fields}, V.~{Pavlidou}, and T.~{Prodanovi{\'c}}, {\it {Cosmic Gamma-ray
  Background from Star-forming Galaxies}},  {\em \apjl} {\bf 722} (Oct., 2010)
  L199--L203, [\href{http://xxx.lanl.gov/abs/1003.3647}{{\tt
  arXiv:1003.3647}}].

\bibitem{2010JCAP...01..005F}
C.-A. {Faucher-Gigu{\`e}re} and A.~{Loeb}, {\it {The pulsar contribution to the
  gamma-ray background}},  {\em \jcap} {\bf 1} (Jan., 2010) 5,
  [\href{http://xxx.lanl.gov/abs/0904.3102}{{\tt arXiv:0904.3102}}].

\bibitem{2012PhRvD..86f3004C}
A.~{Cuoco}, E.~{Komatsu}, and J.~M. {Siegal-Gaskins}, {\it {Joint anisotropy
  and source count constraints on the contribution of blazars to the diffuse
  gamma-ray background}},  {\em \prd} {\bf 86} (Sept., 2012) 063004,
  [\href{http://xxx.lanl.gov/abs/1202.5309}{{\tt arXiv:1202.5309}}].

\bibitem{2012arXiv1206.4734H}
J.~P. {Harding} and K.~N. {Abazajian}, {\it {Models of the Contribution of
  Blazars to the Anisotropy of the Extragalactic Diffuse Gamma-ray
  Background}},  {\em ArXiv e-prints} (June, 2012)
  [\href{http://xxx.lanl.gov/abs/1206.4734}{{\tt arXiv:1206.4734}}].

\bibitem{2010ApJ...712..147A}
A.~A. {Abdo} {\em et.~al.}, {\it {Observations of Milky Way Dwarf Spheroidal
  Galaxies with the Fermi-Large Area Telescope Detector and Constraints on Dark
  Matter Models}},  {\em \apj} {\bf 712} (Mar., 2010) 147--158,
  [\href{http://xxx.lanl.gov/abs/1001.4531}{{\tt arXiv:1001.4531}}].

\end{thebibliography}\endgroup
\bibliographystyle{JHEP}

\end{document}